\documentclass[12pt]{article}
\usepackage{epsfig,graphicx}

\def\be{\begin{equation}}
\def\ee{\end{equation}}
\def\bea{\begin{eqnarray}}
\def\eea{\end{eqnarray}}
\def\ba{\begin{array}}
\def\ea{\end{array}}
\def\re#1{(\ref{#1})}

\newcounter{fig}
\newcounter{tab}

\begin{document}

\begin{titlepage}
\begin{flushright}
%\textbf{May 20, 2003} \\
%\textbf{padsecond9.tex} \\
\end{flushright}
\vskip 2 true cm
%\title{\vspace*{24pt}
\centerline{
{\LARGE\bf {Detailed Empirical Study of the}}
     }
\vskip 0.6 true cm
\centerline{
{\LARGE\bf {Term Structure of Interest Rates. }}
     }
\vskip 0.6 true cm
\centerline{
{\LARGE\bf {Emergence of Power Laws and Scaling Laws }}
     }\vskip 1 true cm
{\large\centerline{\bf Thomas Alderweireld{\small{\footnote[1]{
Thomas.Alderweireld@umh.ac.be,
Universit\'e de Mons-Hainaut,
20 Place du Parc, 7000 Mons, Belgium}}}
and Jean Nuyts{\small{\footnote[2]{Jean.Nuyts@umh.ac.be,
Universit\'e de Mons-Hainaut,
20 Place du Parc, 7000 Mons, Belgium}}}
}}

\vskip 2 true cm
{\centerline{\bf{Abstract}}}
\vskip .5 true cm
{\small
The technique of Pad\'e Approximants, introduced in a previous work, is applied to 
extended recent  data on the distribution of variations of interest rates compiled by the Federal Reserve System
in the US. It is shown that new power laws and new scaling laws emerge for any maturity not only as a function
of the Lag but also as a function of the average inital rate. This is especially true for the one year maturity  where
critical forms and critical exponents are obtained. This suggests future work in the direction of constructing a theory 
of variations of interest rates at a more ``microscopic'' level.}

\end{titlepage}
\eject
\setcounter{page}{1}

\section{Introduction}\label{sec:1}

Many models have been constructed to try to describe and/or predict the term
structure of interest rates. 
Recently, it has been shown on a specific set of data that the empirical
distribution of the variation of interest rates is rather well reproduced
by using the technique of Pad\'e Approximants \cite{NP}. This article is a
further step in this direction. The classical approach is based on Brownian motions
of various kinds. It leads to exponential tails (or superposition of exponential
tails) in the distributions. These so called narrow tails are usually badly
rejected by the data. Levy-flight \cite{M1}, \cite{M2}, \cite{WWW} distributions
have also
been proposed. In this case the tails of the distributions decrease as the
inverse of a small power of the variation of the interest rates
(see below \re{powerd}), in fact a power comprised between one and two. The
predicted tails are
really too fat and are  also rejected by the data. Sometimes rather unjustified
cuts
have been used to sharpen the Levy tails.

In a recent paper \cite{NP}, analyzing numerically the data (Data \{1\}) on the
American daily spot interest rates
compounded by the Federal Reserve System \cite{FRS} between February 15, 1977
and August 4, 1997
($N_{\rm{events}}=5108$ corresponding to 5108 opening days), it was discovered
that the fat tails, known to appear in the
empirical distributions, decrease essentially as a fourth power of the variation
of the interest rate and are thus naturally amenable to ratios of polynomials.
It was shown that the empirical distributions, averaged on the initial interest
rates, are very well fitted with simple Pad\'e Approximants. More recently,
articles taking into account the fat, but not too fat, tails and power laws have
been proposed \cite{DA}-\cite{Bou}.

In this paper, using more complete data (Data \{2\}) recently available, again
from the FRS between January 2, 1962 and December 27, 2002
(10230 opening days, see \re{weight1} below), we have been able to study in a
more detailed way the
Pad\'e fits. At a first stage, we confirmed the preceding analysis. The results
are summarized in section 2.
In section 3, we have refined the analysis by dividing the new data in suitable
subsets, restricting ourselves to intervals in the initial interest rate. This
has allowed us to confirm and to extend the validity of the preceding results,
but also and more importantly to suggest new scaling laws.

These results will be used in a forthcoming article \cite{AN} to try to build a
``microscopic model'' of interest rates.

\section{Pad\'e Approximants fits. Update of the Previous Results}\label{sec:2}

In this section, we present briefly the notation and update the results, which
have been obtained in \cite{NP} for the Data \{1\}, using the full data set (Data \{2\}) 
available today~\cite{FRS}.

Suppose that, at some initial time $t_0$, the interest rate (for some maturity
$[m]$) is $I_0$ and, at some later time $t_f$, it is $I_f$.
The lag and the variation of interest rates were defined as follows
\begin{itemize}

\item
The lag is the time interval between the initial and final times
\be
L=t_f-t_0 \ .
\label{lag}
\ee
It should be stressed that the lag is expressed in consecutive opening days and
that non working days were simply discarded~\cite{Hull}.

\item
The variation of interest rates is
\be
v=I_f-I_0 \ .
\label{varI}
\ee
Since, in the data, the interest rates are expressed exactly in basis points
(0.01 percent), the basis point has been chosen as the natural unit. When
discretized, the variables $I$ and $v$ are thus represented by
integers, $\widehat I$ and $\widehat v$.

\end{itemize}

Let us denote by $p^{[m]}(L,v,I_0)$ the probability distribution
and by $\widehat p^{[m]}(L,\widehat v,\widehat{I_0})$ the corresponding
discretized distribution.
They are the probability that the interest rate for
maturity $[m]$ has moved from $I_0$ to $I_f$ during the lag $L$. Their norms are
\bea
\int_{-\infty}^{\infty}p^{[m]}(L,v,I_0)\,dv\ &=&\ 1 \ ,
    \label{pnormcont}     \\
\ \sum_{\widehat v=-\infty}^{\infty}
\widehat p^{[m]}(L,\widehat v,\widehat{I_0})\ &=&\ 1\ .
\label{pnorm}
\eea

In the FRS Data \{1\} and \{2\}, the maturities
$[m]=$ 1, 2, 3, 5, 7, 10 and  30 years are given.
All the daily interest rates are expressed exactly in basis points. They
vary roughly from $1.5\ \%=150$ basis
points to $18\ \%=1800$ basis points. The discrete empirical distributions
$\widehat w(\widehat I)$ of the initial
interest rates, for a given maturity and for the two sets, are defined by
\bea
\widehat w(\widehat I)&=&\frac{N^d(\widehat I)} 
      {N_{\rm{events}}} \ ,
\label{weight}\\
N^d(\widehat I)&=&
  {\rm{number\ of\ days\ the\ interest\ rate,\ for\ maturity\ }} [m],
        \nonumber\\
&&{\rm{ was\ }}\widehat I
        {\rm{\ basis\ points\ in\ the\ FRS\ Data\ \{1\}\ or \ \{2\}}} \ ,
        \nonumber\\
N_{\rm{events}}&=&5108 \ {\rm{in\ Data\ \{1\}\ for\ all\ }}[m] \ ,
        \nonumber\\
&&\hskip -2.3 true cm
\begin{array}
{|c|c|c|c|c|c|c|c|c}
\cline{1-8}
\ \ \left[m\right]\ \ \  = & 1 & 2 & 3 & 5 & 7 & 10 & 30 &\ {\rm{in\ Data\
\{2\}}} \ . \\
\cline{1-8}
N_{\rm{events}}\ \,=&10230 &6606 &10199 &10199 &8323 &10199 &6214 &\\
\cline{1-8}
\end{array}
\label{weight1}
\eea
As an example,
the FRS empirical distribution $\widehat w$ for the maturity $[m]=$ 1 year and for
the Data \{2\} is given in Figure~\re{wfig}. Obviously, we have for any such distribution
\be
\sum_{\widehat I=150}^{1800} \widehat w(\widehat I)=1 \ .
\label{wsum}
\ee

The main subject of study in  \cite{NP} was in fact the discrete average
$\widehat p_{\rm{average}}^{[m]}(L,\widehat{v})$ defined by
\be
\widehat p_{\rm{average}}^{[m]}(L,\widehat v)=
             \sum_{\widehat{I_0}=150}^{1800} \widehat w^{[m]}(I_0)\
             \widehat p^{[m]}(L,\widehat v,\widehat{I_0})\ .
\label{averp}
\ee
The discretized averages can easily be extracted from the data for
lags (in days) such that $1\leq L\leq 30$ and for variations of interest rates
$\widehat v$ (expressed in basis points) included in the range
$R_v=[-200\leq \widehat{v}\leq 200]$. Indeed, outside this range, the data is
always zero.
The discrete normalization becomes
\be
\sum_{\widehat v=-200}^{200}\widehat p_{\rm{average}}^{[m]}(L,\widehat v)=1 \ .
\ee

It has been first shown empirically, using Data \{1\},
that the tail of the term structure of the
interest rates decreases essentially as a fourth power of the variation of the
interest rates i.e.
\be
p_{\rm{average}}^{[m]}(L,v)\approx \, \kappa
    \, \frac{1}{\mid {\hskip -0.1 cm} v {\hskip -0.1 cm}\mid ^d}
\quad {\rm{when}}\ \mid{\hskip -0.1 cm} v{\hskip -0.1 cm}\mid {\rm{\ becomes\
large}}
\label{powerd}
\ee
with a suitable constant $\kappa$ and a power $d$
\be
d\approx 4 \ .
\label{dfour}
\ee
This means that the distributions decrease much faster than what Levy flight
models  suggest (namely $1\leq d\leq 2$) but much more slowly than what Brownian models
$p_{\rm{average}}\approx \kappa \exp(-v^2/\sigma^2)$ predict.

Using detailed fits, it has been shown in \cite{NP} that the average
distribution \re{averp} can be very well (amazingly well)
approximated by a Pad\'e Approximant $P(0,4)$, i.e. with a constant
numerator and a polynomial of fourth degree in the denominator. More precisely,
for all maturities $[m]$ and for all lags $L$, the following Pad\'e form of the
probability distributions, in terms of $\widehat{v}$,
\be
\widehat p_{\rm{average}}^{[m]}(L,\widehat v)=\frac{q_1}
{\pi\Bigl(1+(q_1^2+2q_2)\widehat v^2+q_2^2\widehat v^4\Bigr)} \ ,
\label{Pade}
\ee
fits the Data \{2\} (and the Data \{1\} set) rather well.

Let us enumerate a few of the properties of this form and of the two remaining
real parameters $q_1$ and $q_2$ which depend on the lag $L$ and on the maturity
$[m]$.

\begin{enumerate}

\item The continuous distribution corresponding to \re{Pade} is, for $q_1$ positive, a purely positive real
function. In fact it is constructed as the modulus square of the complex
function $ f$
\bea
 f_{\rm{average}}^{[m]}(L, v)
&=&\frac{\sqrt{q_1}}{\sqrt{\pi}\left(1+iq_1  v+q_2 v^2\right)} \ ,
             \label{sqrtfunc} \\
 p_{\rm{average}}^{[m]}(L, v)
&=&\mid f_{\rm{average}}^{[m]}(L, v)\mid^2\ .
\label{sqrtprob}
\eea

\item The normalization \re{pnormcont} of the distribution \re{sqrtprob} is
automatic by construction.

\item The distribution depends only on
$v^2$, hence it is symmetrical
under the
interchange $\{v\leftrightarrow -v\}$. This symmetry was checked empirically to
hold to a very precise degree. In other words, violations of this symmetry are
zero well within one standard deviation.

\item The tail of the distribution \re{Pade} is of the form $v^{-4}$. The
Bayesian Hill estimators \cite{Hill} of the empirical tails favor an
approximate decrease of this kind.

\item An estimate of $q_1$ is given by 
\be
q_1=\pi \, \widehat p_{\rm{average}}^{[m]}(L,0)
\label{maxim}
\ee
\noindent
and of $q_2$ by 
\be
q_2 = -\sigma^{-1} 
\label{variance}
\ee
\noindent 
where $\sigma$ stands for the variance of the data distribution.

\item The parameters $q_1$ and $q_2$ have very simple regular forms in term of
the lag $L$ when $L$ varies from one day to about a month and for any maturity
$[m]$. Essentially 
\be
q_i\,\approx\,\beta_i\,\Biggl({\frac{L}{L_{\rm{r}}}}\Biggr)^{-\mu_i\,}    \ \ , \quad i=1,2
\label{qi}
\ee
\noindent
where, for convenience, we have chosen one arbitrary reference value $L_{\rm{r}}=15{\rm{\ days}}$.
The $\mu_i$, which we call scaling parameters, and the less essential multiplication 
constants $\beta_i$ depend on $[m]$ but not on $L$. The Table~\re{mui} (resp. Table~\re{muiold}) 
shows the $\mu_i$ and the ratio ${\mu_2}/{\mu_1}$ that we have obtained for the 7 investigated maturities using the Data \{2\} 
(resp. Data \{1\}) set.

In other words, as illustrated by the Figures~\re{L2q1} and~\re{L2q2} for $[m]=$ 2 years, as an example, in a $\ln(q_i)$ versus $\ln(L)$ 
plot the data points align along a straight line with slope $-\mu_i$.
The plots for the other $[m]$'s follow analogous straight lines.

\item Moreover, one may note that one critical exponent is approximatively related to the other
exponent by (see Tables~\re{mui}~\re{muiold})
\be
\mu_2\approx 1.8\ \ \mu_1 \ ,
\label{scal1exp}
\ee

\item
It should be stressed that this scaling law \re{qi} seems to be an important discovery~\cite{NP}
 which had escaped attention.

\end{enumerate}

\section{Pad\'e Approximants fits. Initial Interest Rate Dependence}\label{sec:3}

More recently, we have studied the  extended set of data (Data \{2\})  and performed
 a more refined analysis.

Indeed, the new data having often a higher statistics, meaningful
subsets of data can be defined with initial interest rates limited to
intervals. We have found that subsets containing about two thousand events are
large enough to lead to a sufficiently precise determination of the parameters.

A detailed numerical analysis of the distributions restricting
the data to an almost fixed initial interest rate $I_0$ has convinced us that
the same Pad\'e behavior prevails. An approximate determination of the
dependence of the Pad\'e
parameters $q_i$ on initial interest rate $I_0$ follows. It suggests that
new scaling laws are at work.

Let us summarize our results

\begin{enumerate}

\item

We have first
divided the Data \{2\} set in subsets where $I_0$ is restricted to regions
(overlapping or not)
\be
\Delta=[\widehat I_{0{\rm{min}}},\widehat I_{0{\rm{max}}}] \ .
\label{regions}
\ee
The total probability $\omega_{\Delta}$ to be inside $\Delta$ and the total numbers of events $N_{\Delta}$ inside $\Delta$
are given by
\bea
\omega_{\Delta} & = & {\sum_{\widehat I=\widehat I_{0{\rm{min}}}}
                ^{\widehat I_{0{\rm{max}}}}
                \widehat w(\widehat I)} \ , \\
N_{\Delta} & = & \omega_{\Delta} N_{\rm events} \ .
\eea
In the present analysis, the regions have be chosen in such a way that  the $N_{\Delta}$'s 
are around two thousands. If the number of events is chosen to be smaller than
about two thousand, the accuracy in the determination of the parameters becomes
hazardous. On the other hand, there seems to be no need to select larger regions
with more events in them. The events in Data \{2\} can thus, for
example for $[m]=$ 1 year, be separated
in five non overlapping intervals covering the full $\widehat I_0$ range.

\item

Let us define $\overline{I_0}$ as the empirical average of $\widehat I$ in a
given region $\Delta$
\be
\overline{I_0}=
       \frac{\sum_{\widehat I=\widehat I_{0{\rm{min}}}}
                ^{\widehat I_{0{\rm{max}}}}
              \widehat w(\widehat I)\,\widehat I}
              {\omega_{\Delta}} \ .
 \label{meanI0}
\ee

\item

The distributions, for the events within a region $\Delta $ (see \re{average} for a precise definition), 
are then supposed, in first approximation, to be represented by normalized \re{pnormcont} Pad\'e forms
\be
\overline{p}(L,\widehat v,\overline{I_0})
   =\frac{q_1}{\pi\bigl(1+(q_1^2+2q_2)\widehat v^2+q_2^2\widehat v^4\bigr)}
     \ .
\label{Padepp}
\ee

\item

The parameters $q_i$ appearing in Eq.\re{Padepp} (for a given maturity $[m]$)
are thus evaluated as
functions of $L$ and $\overline{I_0}$. They are obtained from the data by
minimizing a $\chi^2$ function.
In about all instances the fits are very good as exemplified by the
$\chi^2$ values.

\item

In Figure~\re{Padefig}, as an example using the Data \{2\} with $L=1$ and
$[m]=$ 1 year, two thousand points in the region
$\Delta=$[420 basis points, 553 basis points] (see~\re{regions}) have been selected and the empirical points 
are plotted versus the best Pad\'e fit. 
The average initial interest rate~\re{meanI0} is
about $\overline{I_0}=502$ basis points. In this case, the values of $q_i$ evaluated in their respective units are
$q_1=38.6 \pm 1.3 (\%)^{-1}$ and $q_2= -308 \pm 19 (\%)^{-2}$. The obtained $\chi^2$ is 61 with 83 degrees of freedom.
In almost all $L,[m]$ cases an eye check of the curves is as satisfactory and the computed $\chi^2$ are as good.

\item

At this point, an important remark is worth making.
It should be stressed that the parameters $q_i$ obtained this way are not really
those at $I_0=\overline{I_0}$ itself but a complicated average of the
approximate $q_i$ in the corresponding $\Delta$ range. Indeed, it is obvious
that the
superposition of two or more Pad\'e forms is, at best, only approximatively of a
Pad\'e form. Since this is an important point, let us explain it more precisely.
If $\widehat p(L,\widehat v, \widehat I)$ is known for any $\widehat I$, then the average probability
$\overline{p}(L,\widehat v,\overline{I_0})$ \re{Padepp} on $\Delta$
is defined by
\be
\overline{p}(L,\widehat v,\overline{I_0})
                      =\frac
                  {\sum_{\widehat I=\widehat I_{0{\rm{min}}}}
                       ^{\widehat I_{0{\rm{max}}}}
           \widehat w(\widehat I  )
                         \, \widehat p(L,\widehat v,\widehat I  )}
                  {\omega_{\Delta}}
               \ .
\label{average}
\ee
If it then were true that $\widehat p(L,\widehat v,\widehat{I})$ is exactly of
the Pad\'e form
(which is anyhow not a correct statement but, at best, an approximate one),
the $\Delta$--average probability $\overline p(L,\widehat v,\overline{I_0})$ would not
exactly be of a Pad\'e form.

\item

After the fits have been performed, we find that the new parameters $q_i$ depend
effectively and importantly on $\overline{I_0}$.

\item

In a first approach, we have chosen to minimize the natural $\chi^2$
leaving $q_1$ as free parameter and by determining $q_2$ from the experimental variance
using the relation~\re{variance}. The interested reader can find the precise definition of the $\chi^2$ in the reference \cite{NP}, in and
after their Eq.(22).

\item We defer to the next section a detailed discussion of the parameters $q_i$, obtained from 
the data, as functions of the lag $L$ and of the average interest rate $\overline{I}_0$.

%The $q_i$'s are given in Figures~\re{q1par2fig} and \re{q2par2fig} for a maturity $[m]=[1]$ year and a lag $L=1$ 
%day as a function of $\overline{I_0}$.

\end{enumerate}

\section{Pad\'e parameters as a function of the lag, of the maturity and, of the initial interest rate}

In this section, we summarize the scaling law and the $\overline{I}_0$ dependence that are suggested by the
empirical data.

\subsection{General considerations and scaling laws}
The first scaling law and the $I_0$ dependence can be summarized as follows. For each
maturity, the empirical $q_i$ are given by the approximate forms (compare to~\re{qi})
\be
q_i\,\approx\,\beta_i\,\Biggl({\frac{L}{L_{\rm{r}}}}\Biggr)^{-\mu_i\,}
           {\mathcal F}_i \Biggl({\frac{\overline{I_0}}{I_r}}\Biggr) \ .
\label{superqi}
\ee
where the ${\mathcal F}_i$ are functions depending of the initial interest rate. For convenience, we have
chosen two arbitrary reference values
\bea
L_{\rm{r}}&=&15{\rm{\ days}}
       \nonumber\\
I_{\rm{r}}&=&600{\rm{\ basis\ points}}\ .
\label{refLI0}
\eea
In \re{superqi}, the lag $L$ is in days, the average initial rate $\overline I_0$
is in basis points, the parameters $q_1$ and $\beta_1$ are in inverse
basis points, the parameters
$q_2$ and $\beta_2$ are in inverse basis points squared while the
scaling parameters $\mu_i$ are pure numbers.

The scaling law \re{superqi}  says that, for every maturity $[m]$, each of the
 parameters $q_i$ is the product of two functions. One function depends on
 the lag $L$ only and has the form of a power law as~\re{qi}. The second
 function, ${\mathcal F}_i$, depends on $\overline I_0$ only.

The best values for the parameters $\mu_i$  are given in Table~\ref{mui} as a function of the maturity.
The form of the  ${\mathcal F}_i$ can be found in the fourteen plots given in 
Figures~\re{q1par2fig}-\re{q2par2figrest}. In Figure~\re{q1par2fig}, for $[m]=$ 1 year and $L=1$ day, the $\ln(q_1)$ is plotted 
versus $\overline I_0/I_{r}$. In Figure~\re{q2par2fig}, for $[m]=$ 1 year and
$L=1$ day, the $\ln(q_2)$ is plotted versus $\overline I_0/I_{r}$.
In Figure~\re{q1par2figrest} there are six plots where for
$L=1$, $\ln(q_1)$ is given versus $\overline I_0/I_r$ for
$[m]=$ 2, 3, 5, 7, 10, 30 years.
Finally, in Figure~\re{q2par2figrest} there are six plots where for
$L=1$, $\ln(q_2)$ is given versus $\overline I_0/I_r$ for
$[m]=$ 2, 3, 5, 7, 10, 30 years.

We have shown that the assumption in~\re{superqi} that the $q_i$ factorize
(separation of variables) is favoured by the data. Hence, for a given maturity, the distributions of  $\ln(q_i)$ as a function 
of $\overline{I}_0$ for different $L$ values just differ by a constant shift. This shift on the $\ln(q_i)$ axis is equal to 
\be
\ln(q_i)-\ln(q_i')=\mu_i \ln(L'/L) \ .
\label{delta}
\ee

To illustrate this, the Figures~\re{deltaq1} and \re{deltaq2} show  the distributions of $\ln(q_i)$, for $[m]=$ 5 years as a 
function of $\overline{I}_0$ for 3 different $L$ values (1, 10 and 30 days). To make the comparison easier, the $L$=10 days and the $L$=30 days distributions 
have been moved upwards by their respective shift term ($\mu_i \ln(10)$ and $\mu_i \ln(30)$  respectively). For the other maturity, one obtains similar plots.

It has to be emphasized that this variable separation in the approximate forms may be a hint  of a scaling law 
which is at work.

\subsection{The scaling laws  for the $[m]=$ 1 year case}
For the $[m]>1$ year data, no particular simple shape for the function  ${\mathcal F}_i$ can be extracted. In the $[m]=$ 1 year 
data, the functions ${\mathcal F}_i$ have a very simple shape. The $\ln(q_i)$  distributions as function of $\overline{I_0}$  (see Figures~\re{q1par2fig}\re{q2par2fig} for $L=1$ day) 
can be fitted  with  straight lines and the following extremely simple approximate forms hold  
\be
q_i\,\approx\,\beta_i\,\Biggl({\frac{L}{L_{\rm{r}}}}\Biggr)^{-\mu_i\,}
          \exp\Biggl( \Biggl({\frac{\overline{I_0}}{I_r}}\Biggr)^{-\nu_i}\Biggr) \ .
\label{superqi1year}
\ee
\noindent
where  
\bea
\ln(\beta_1) = 3.14 \pm 0.02, \ \ \mu_1 = 0.657 \pm 0.005, \ \ \nu_1 = 1.32 \pm 0.02  \\
\ln(\beta_2) = 4.51 \pm 0.02, \ \ \mu_2 = 1.197 \pm 0.005, \ \ \nu_2 = 2.19 \pm 0.02 \ .
\eea
Hence, for the particular case of $[m]=$ 1 year, we obtain an interplay of
two simple scaling laws with critical exponents $\mu_i$ and $\nu_i$.

\section{Conclusion}

In this paper, we have extended an earlier work aimed at approximating the term
structure of interest rates by ratios of polynomials called Pad\'e Approximants.
The form which is used is a
purely positive Pad\'e $P[0,4]$ with a constant numerator and a fourth degree
denominator in the variation of interest rates.

We have shown that for any maturity, for any lag and for initial interest rates
restricted to regions, the form holds to a very good degree of approximation,
better than what should have been hoped a priori.

The empirical parameters are represented by functions which point toward the
existence of scaling laws and scaling exponents at a more ``microscopic level''.
These scaling laws were already discovered and discussed \cite{NP} in terms the
lag variable. In this paper, using the extended FRS data \cite{FRS} now
available, we have shown that scaling laws seem also to be at work not
only in the lag, as was shown previously, but also in the initial interest rate.
This is true especially for $[m]=$ 1 year maturity where simple critical forms \re{superqi1year} appear both
in the lag $L$ and in the average interest rate $\overline{I}_0$ and critical exponents can be extracted. 

In a forthcoming paper, we will try to use this result and the scaling laws 
which are suggested by the data to build a theory of the term structure of interest 
rates at a ``microscopic level''.

\vskip .7true cm
\noindent{\large\bf{Acknowledgment}}
\vskip 0.2true cm
\noindent

The work of
one author (T.A.) was supported in part by Belgian
I.I.S.N. (Institut Interuniversitaire des Sciences Nucl\'eaires).
The work of the other (J.N.) was supported in part by
the Belgian F.N.R.S. (Fonds National de la Recherche Scientifique). He would
also like to thank Pr.$\,$Isabelle Platten for discussions and numerical simulations
in an early phase of this research.

\newpage
\centerline{\bf{Figure Caption}}

\begin{description}
\refstepcounter{fig}

\vskip 0.5 true cm
\item{Figure \thefig.}
{\label{wfig}}
\refstepcounter{fig}

The FRS empirical weight distribution  $\widehat w$ of the interest rates (see \re{weight}) for the maturity
$[m]=$ 1 year and for the Data set \{2\}.

\vskip 0.5 true cm
\item{Figure \thefig. }
{\label{L2q1}}
\refstepcounter{fig}

Distribution of $\ln(q_1)$ as a function of $\ln(L/15)$ for $[m]=$ 2  years using  Data set \{2\}.

\vskip 0.5 true cm
\item{Figure \thefig.}
{\label{L2q2}}
\refstepcounter{fig}

 Distribution of $\ln(q_2)$ as a function of $\ln(L/15)$  for $[m]=$ 2 years using  Data set \{2\}.

\vskip 0.5 true cm
\item{Figure \thefig. }
{\label{Padefig}}
\refstepcounter{fig}

One example of a fit to the  Data set \{2\}  for $[m]=$ 1 year and $L= 1$ day by a Pad\'e Approximant (continuous curve). 
The chosen subset (see \re{regions}) corresponds to 2000 points in the region $I_{0{\rm{min}}}=420$ and $I_{0{\rm{max}}}=553$. 
The $\overline{I}_0$ is computed to be $502$ basis points. The empirical points are plotted versus the best Pad\'e fit. 
In this case, the values of $q_i$ evaluated in their respective units are $q_1=38.6 \pm 1.3 (\%)^{-1}$ and $q_2= -308 \pm 19 (\%)^{-2}$. 
The obtained $\chi^2$ is 61 with 83 degrees of freedom.

\vskip 0.5 true cm
\item{Figure \thefig.}
{\label{q1par2fig}}
\refstepcounter{fig}

The parameter $\ln(q_1)$ as a function of ${\overline{I_0}}/{600}$ for $[m]=$ 1 year
and $L= 1$ day. These points have been obtained by minimizing the $\chi^2$ as a function 
of parameter $q_1$. Each point corresponds to a subset of data containing close to two thousand points 
with average $I_0$ value equal to $\overline{I_0}$. The subsets are overlapping so that the points are not
independent. The unit used for $q_1$ is $(\%)^{-1}$. For $\overline{I_0}$ it is the basis point.
The straight line is a linear fit. Its equation is $\ln(q_1)=\ln(\beta_1) -\nu_1({\overline{I_0}}/{600}) - \mu_1 \ln(1/15)$ with
$\nu_1= 1.32 \pm 0.02 $ and $\ln(\beta_1) = 3.14 \pm 0.02$.

\vskip 0.5 true cm
\item{Figure \thefig.}
{\label{q2par2fig}}
\refstepcounter{fig}
\vskip 0.2 true cm

The parameter $\ln(q_2)$ as a function of ${\overline{I_0}}/{600}$ for $[m]=$ 1 year
and $L= 1$ day. These points have been obtained using the relation defined in \re{variance}. 
Each point corresponds to a subset of data containing close to two thousand points 
with average $I_0$ value equal to $\overline{I_0}$. The subsets are overlapping so that the points are not
independent. The error on $\ln(q_2)$ is computed by propagating the error on the estimate of the variance.  
The unit used for $q_2$ is $(\%)^{-2}$. For $\overline{I_0}$ it is the basis point.
The straight line is a linear fit. Its equation is $\ln(q_2)=\ln(\beta_2) -\nu_2({\overline{I_0}}/{600}) - \mu_2 \ln(1/15)$
with $\nu_2= 2.19 \pm 0.02 $ and $\ln(\beta_2)= 4.51 \pm 0.02$.

\vskip 0.5 true cm
\item{Figure \thefig.}
{\label{q1par2figrest}}
\refstepcounter{fig}

The parameter $\ln(q_1)$ as a function of ${\overline{I_0}}/{600}$ for $[m]=$ 2, 3, 5, 7, 10, 30 years
and $L= 1$ day. These points have been obtained by minimizing the $\chi^2$ as a function 
of parameter $q_1$. Each point corresponds to a subset of data containing close to two thousand points 
with average $I_0$ value equal to $\overline{I_0}$. The subsets are overlapping so that the points are not
independent. The unit used for $q_1$ is $(\%)^{-1}$. For $\overline{I_0}$ it is the basis point.

\vskip 0.5 true cm
\item{Figure \thefig.}
{\label{q2par2figrest}}
\refstepcounter{fig}

The parameter $\ln(q_2)$ as a function of ${\overline{I_0}}/{600}$ for $[m]=$ 2, 3, 5, 7, 10, 30 years
and $L= 1$ day. These points have been obtained using the relation \re{variance}. 
Each point corresponds to a subset of data containing close to two thousand points 
with average $I_0$ value equal to $\overline{I_0}$. The subsets are overlapping so that the points are not
independent. The unit used for $q_2$ is $(\%)^{-2}$. For $\overline{I_0}$ it is the basis point.
The error on  $\ln(q_2)$ is computed by propagating the error on the estimate of the variance.

\vskip 0.5 true cm
\item{Figure \thefig.}
{\label{deltaq1}}
\refstepcounter{fig}

Distribution of $\ln(q_1)$, for $[m]=$ 5 years as a  function of  ${\overline{I_0}}/{600}$ for 3 different $L$ values (1,10 and 30 days)  
where the $L$= 10 and 30 days distributions have been shifted  on the  $\ln(q_1)$ axis by the correcting term \re{delta}.

\vskip 0.5 true cm
\item{Figure \thefig.}
{\label{deltaq2}}
\refstepcounter{fig}

Distribution of $\ln(q_2)$, for $[m]=$ 5 years as a  function of ${\overline{I_0}}/{600}$ for 3 different $L$ values (1,10 and 30 days)  
where the $L$= 10 and 30 days distributions have been shifted  on the  $\ln(q_2)$ axis by the correcting term \re{delta}.

\end{description}

\centerline{\bf{Table Caption}}

\refstepcounter{tab}

\begin{description}

\vskip 0.5 true cm
\item{Table \thetab. }
{\label{mui}}
\refstepcounter{tab}

The values of the best fits to the parameters $\mu_i$ (see \re{superqi}) using  Data set~\{2\}. 
These parameters are given as a function of the maturity $[m]$. 
The standard errors correspond to a unit deviation in the $\chi^2$. 

\vskip 0.5 true cm
\item{Table \thetab.}
{\label{muiold}}
\refstepcounter{tab}

The values of the best fits to the parameters $\mu_i$ (see \re{superqi}) using  Data set~\{1\}.
These parameters are given as a function of the maturity $[m]$. 
The standard errors correspond to a unit deviation in the $\chi^2$. 

\end{description}
\newpage
\centerline{\bf{Figure \re{wfig}}}
\begin{center}
\epsfxsize=8.5cm\epsffile{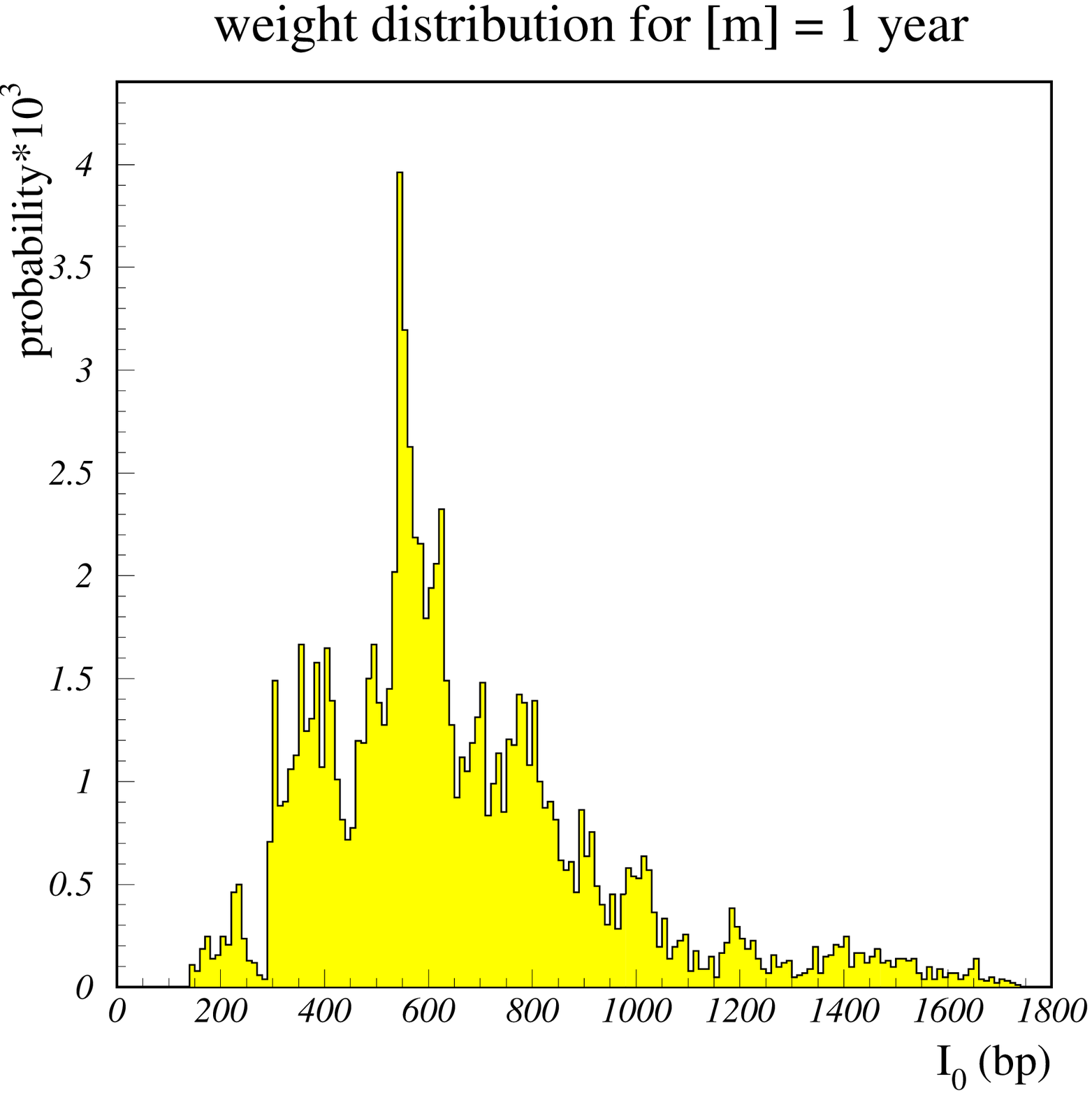}
\end{center}

\centerline{\bf{Figure \re{L2q1}}}
\begin{center}
\epsfxsize=8.5cm\epsffile{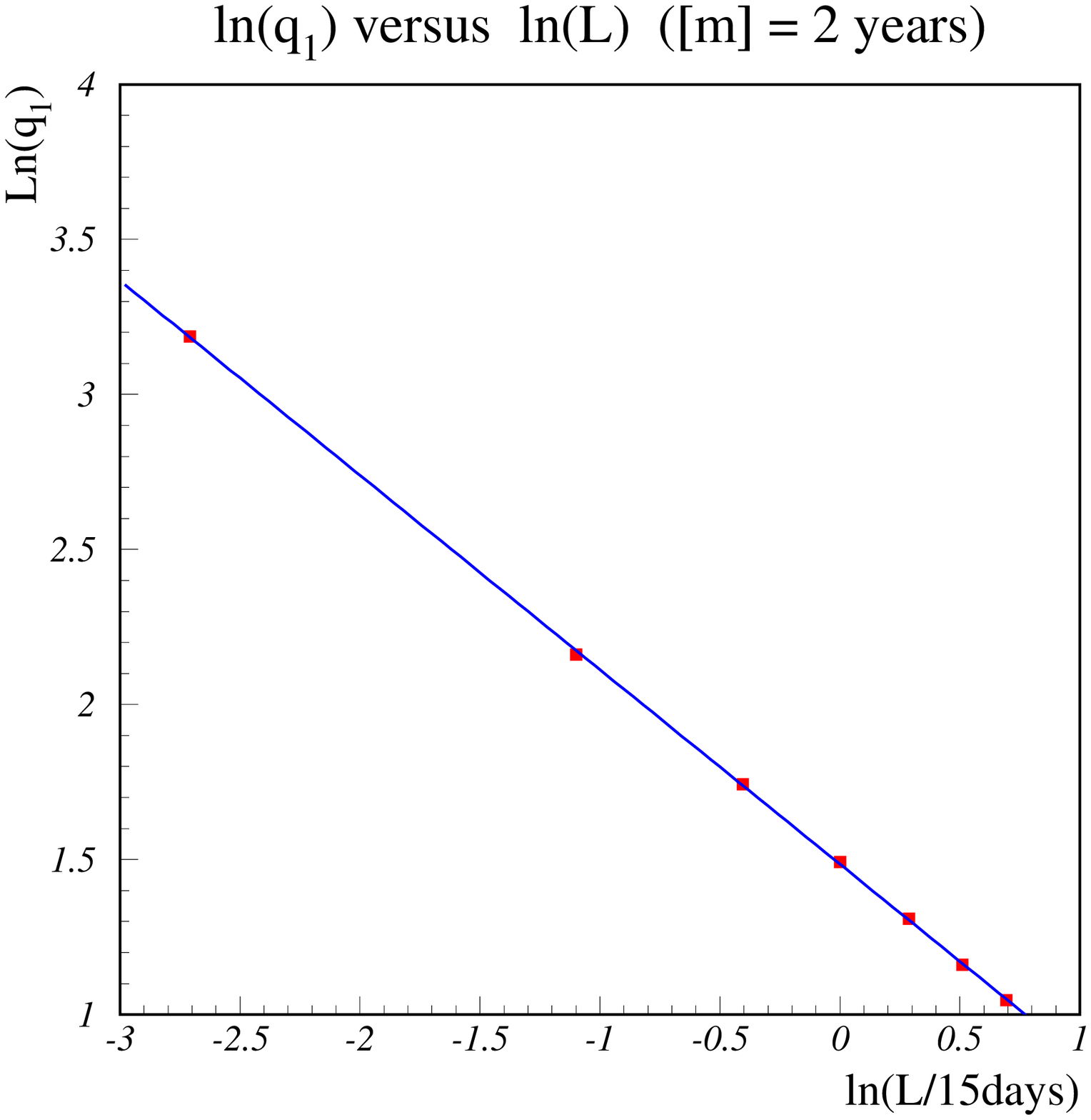}
\end{center}

\newpage
\centerline{\bf{Figure \re{L2q2}}}
\begin{center}
\epsfxsize=8.5cm\epsffile{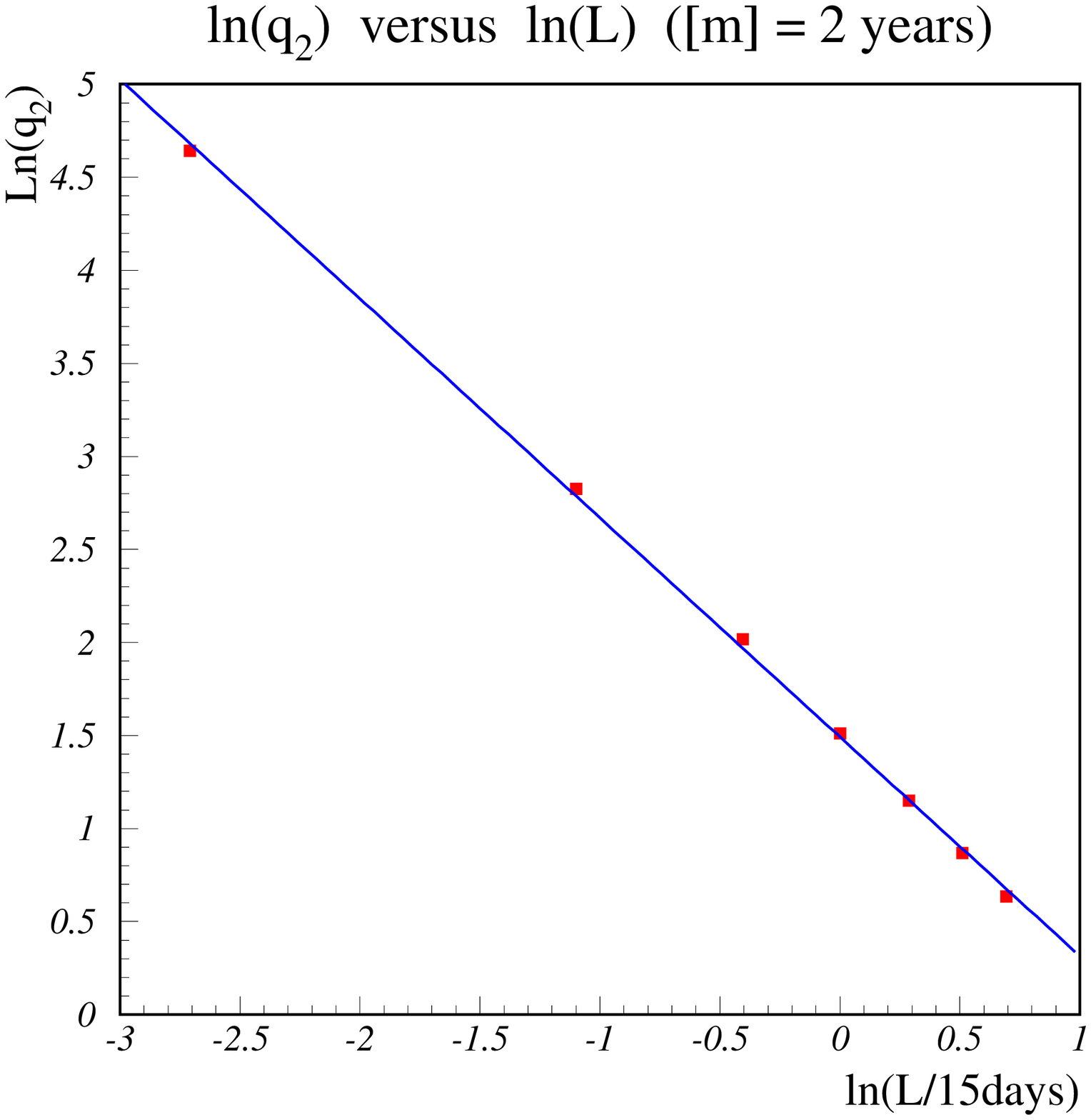}
\end{center}

\centerline{\bf{Figure \re{Padefig}}}
\begin{center}
\epsfxsize=8.5cm\epsffile{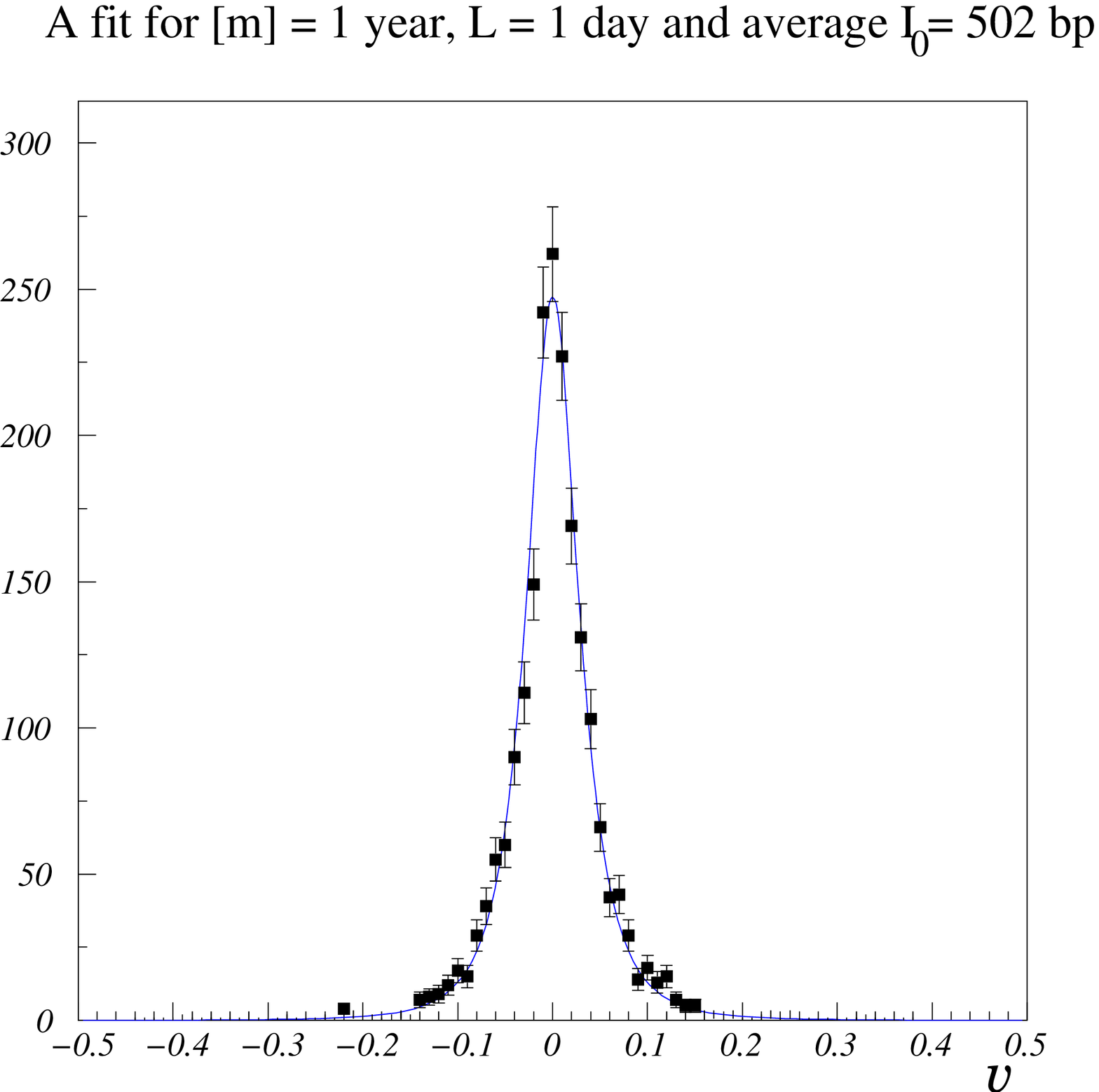}
\end{center}

\newpage

\centerline{\bf{Figure \re{q1par2fig}}}
\begin{center}
\epsfxsize=8.5cm\epsffile{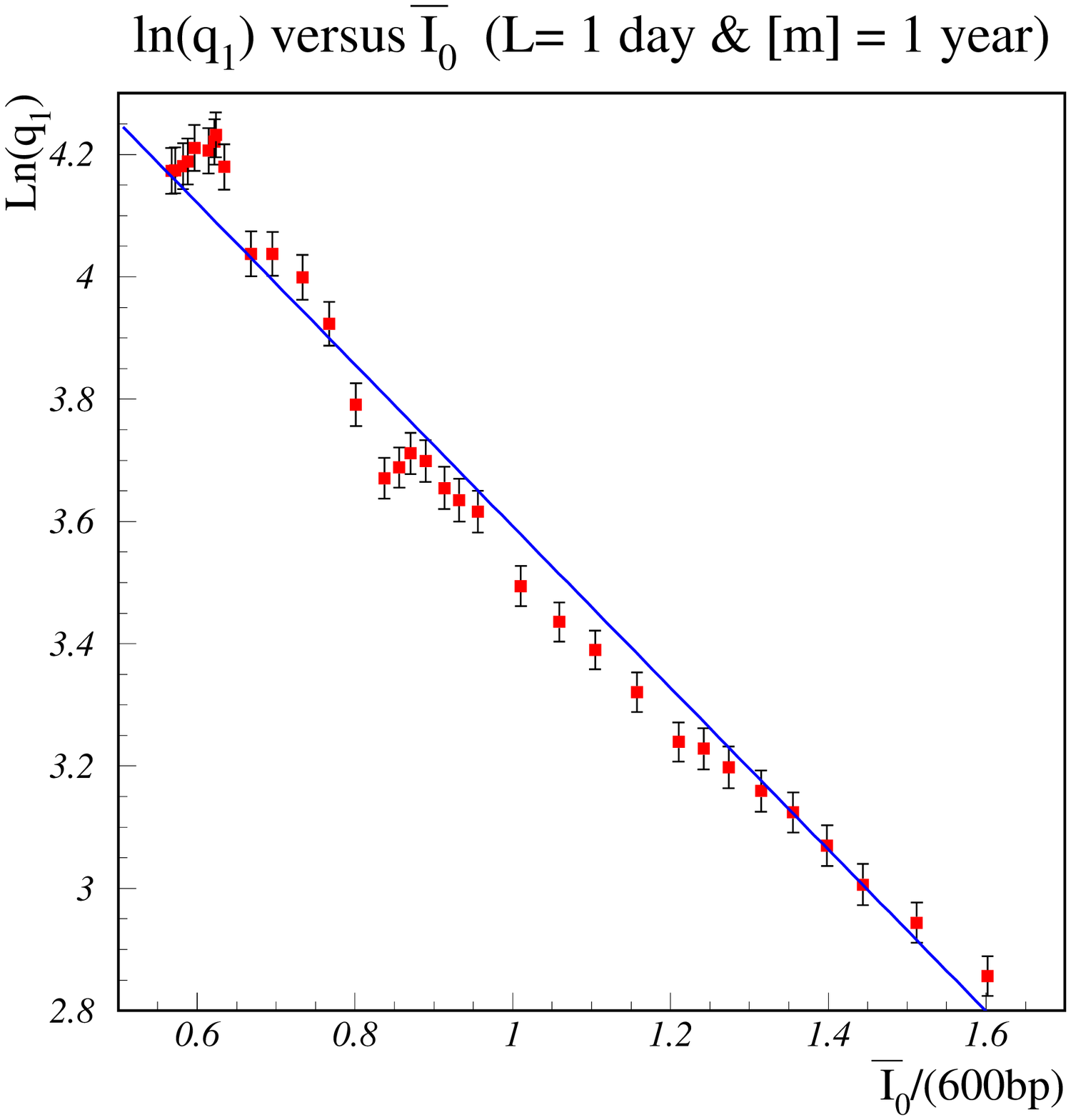}
\end{center}

\centerline{\bf{Figure \re{q2par2fig}}}
\begin{center}
\epsfxsize=8.5cm\epsffile{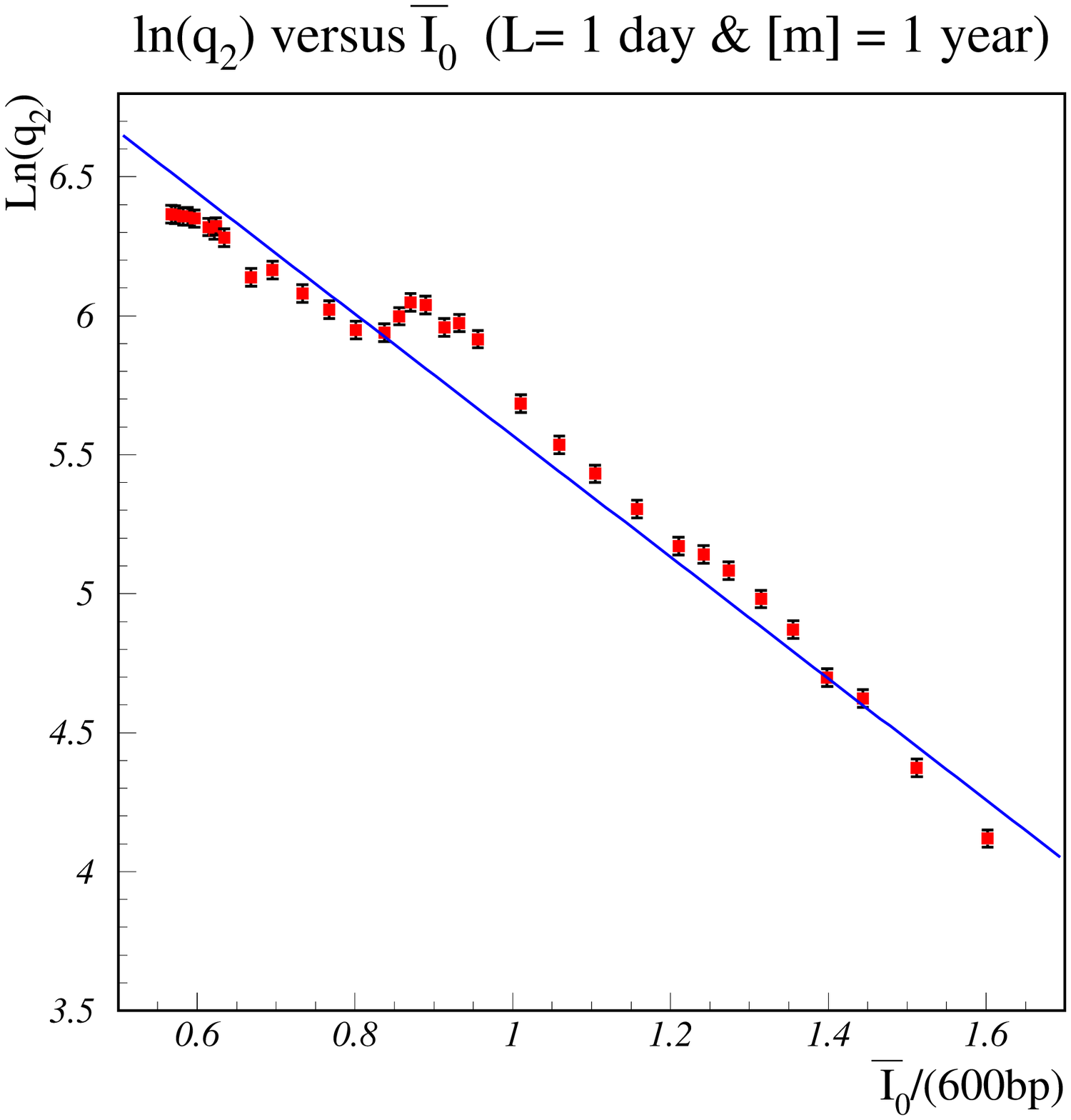}
\end{center}

\newpage

\centerline{\bf{Figure \re{q1par2figrest}}}
\vspace{-0.7cm}
\begin{center}
\begin{tabular}{cc}
 \epsfxsize=6.4cm  
 \epsffile{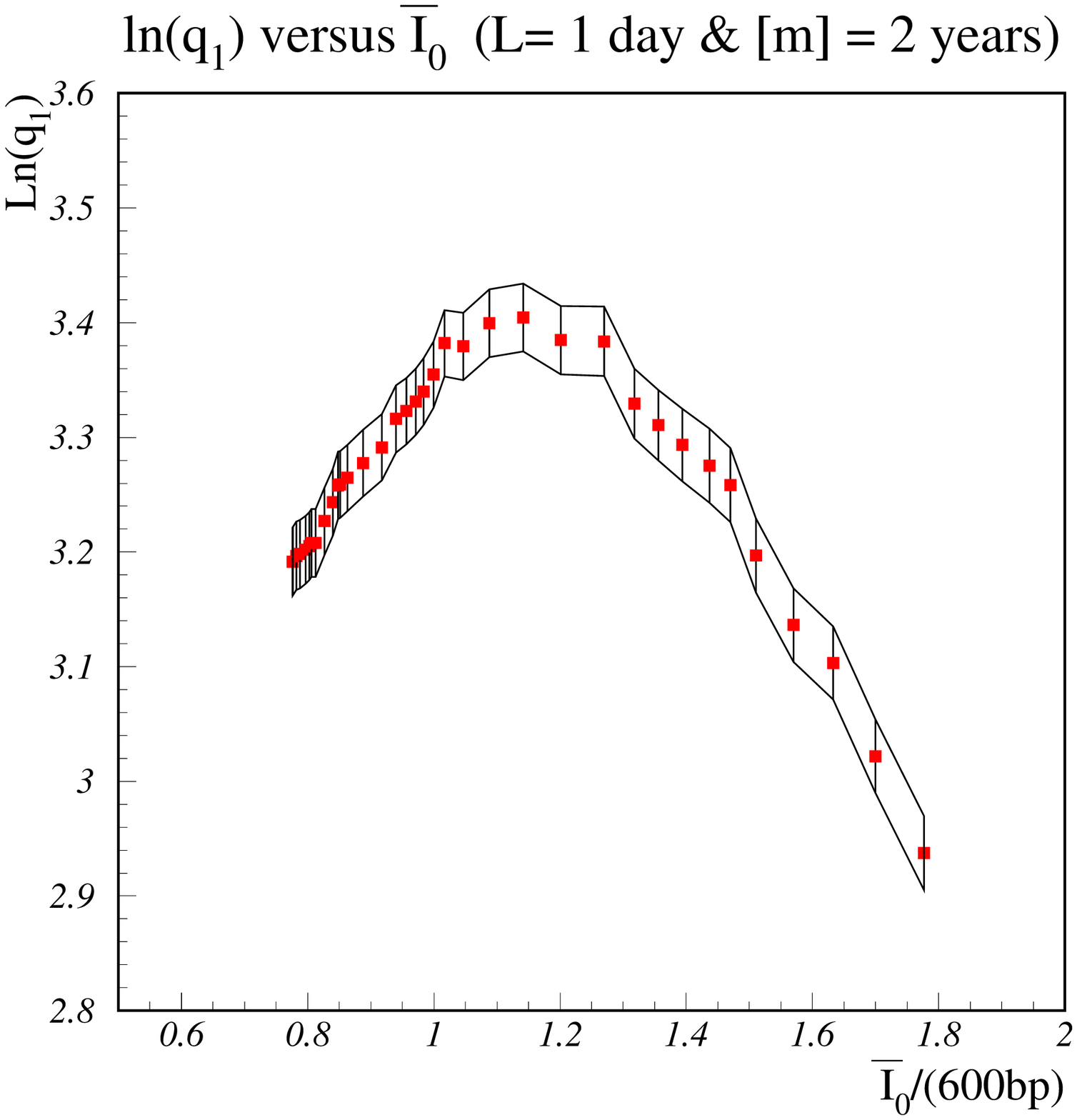}
 &
 \epsfxsize=6.4cm  
 \epsffile{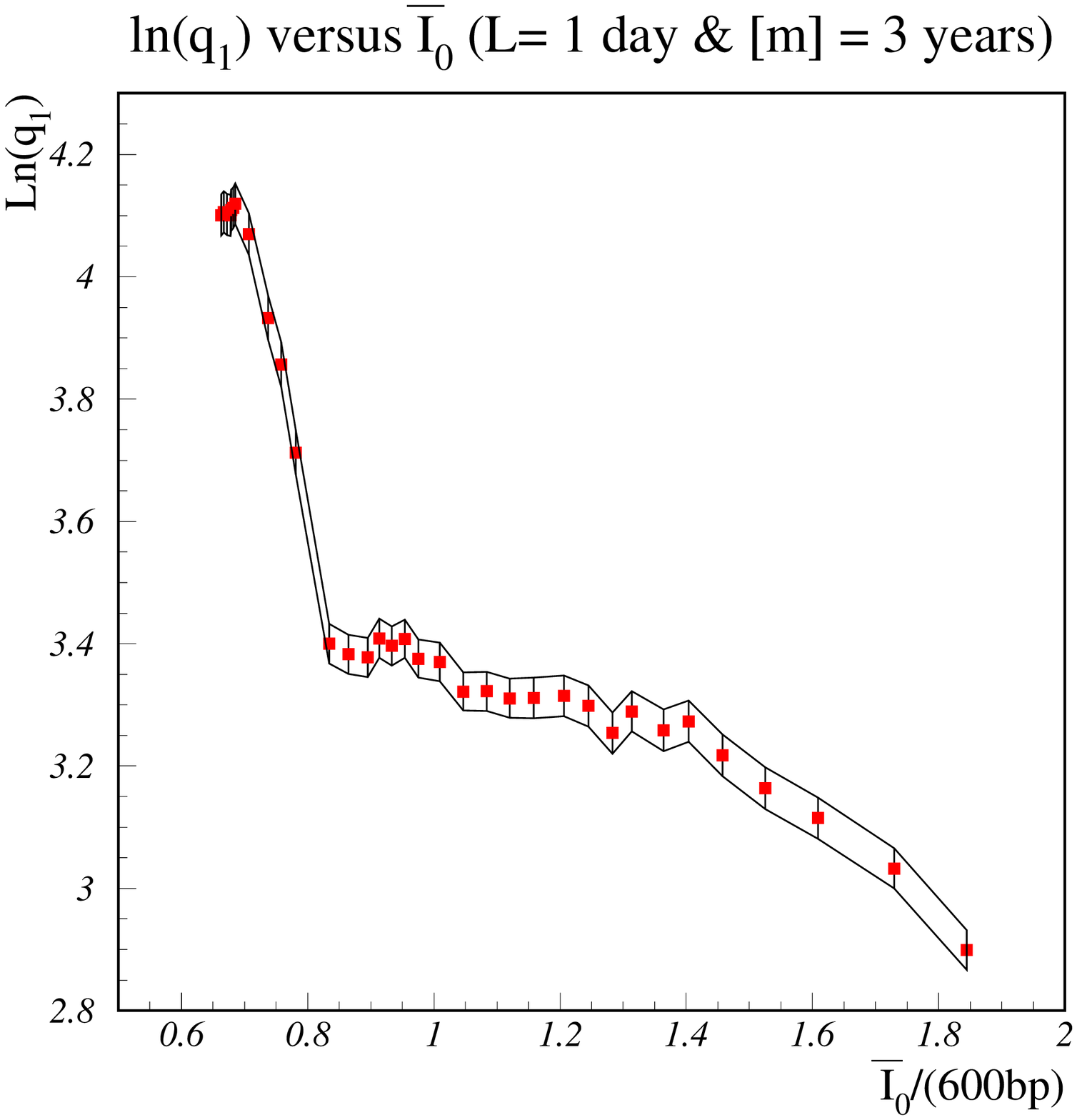}
 \\
 \epsfxsize=6.4cm  
 \epsffile{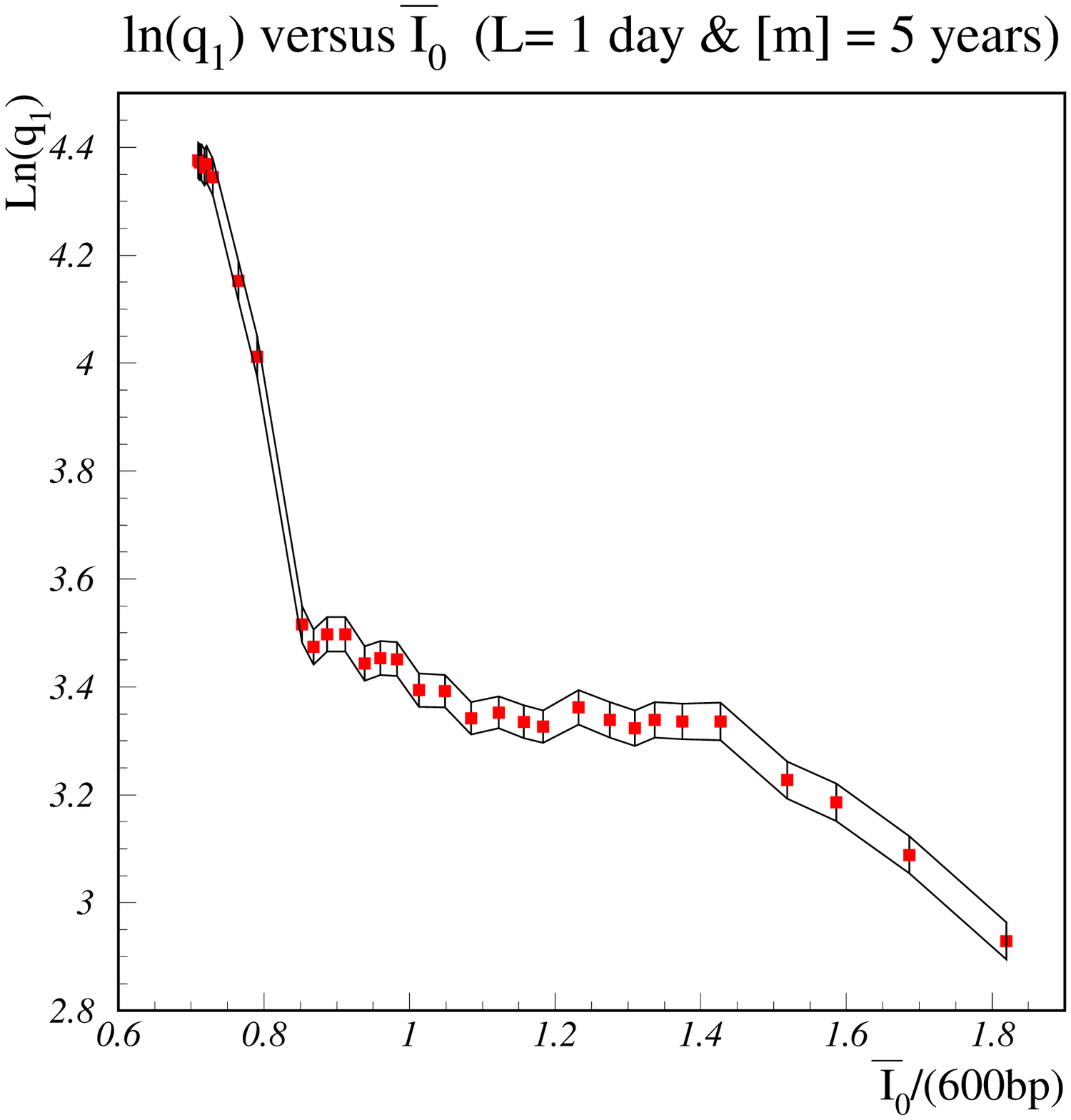}
 &
 \epsfxsize=6.4cm  
 \epsffile{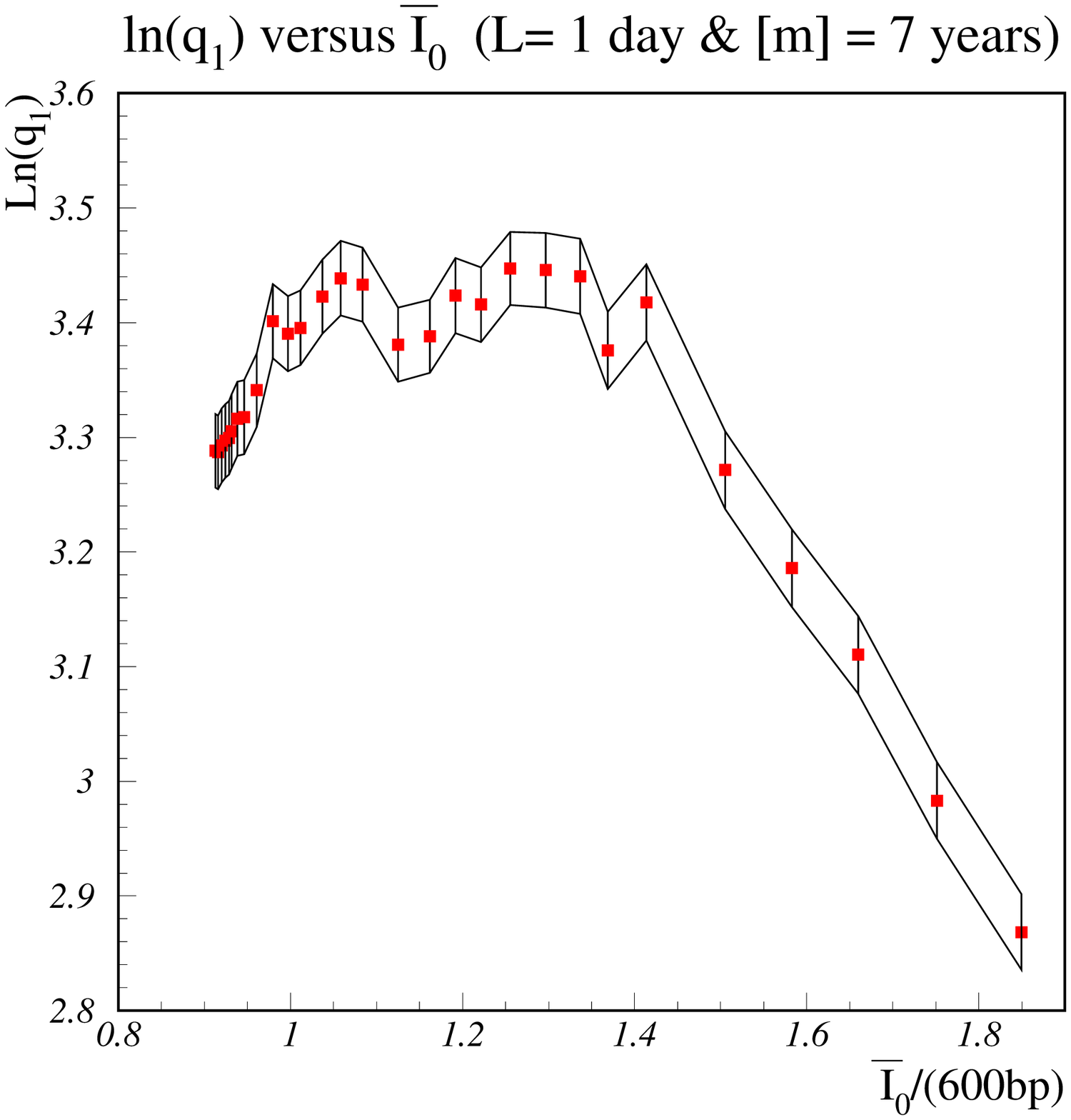}
 \\
 \epsfxsize=6.4cm  
 \epsffile{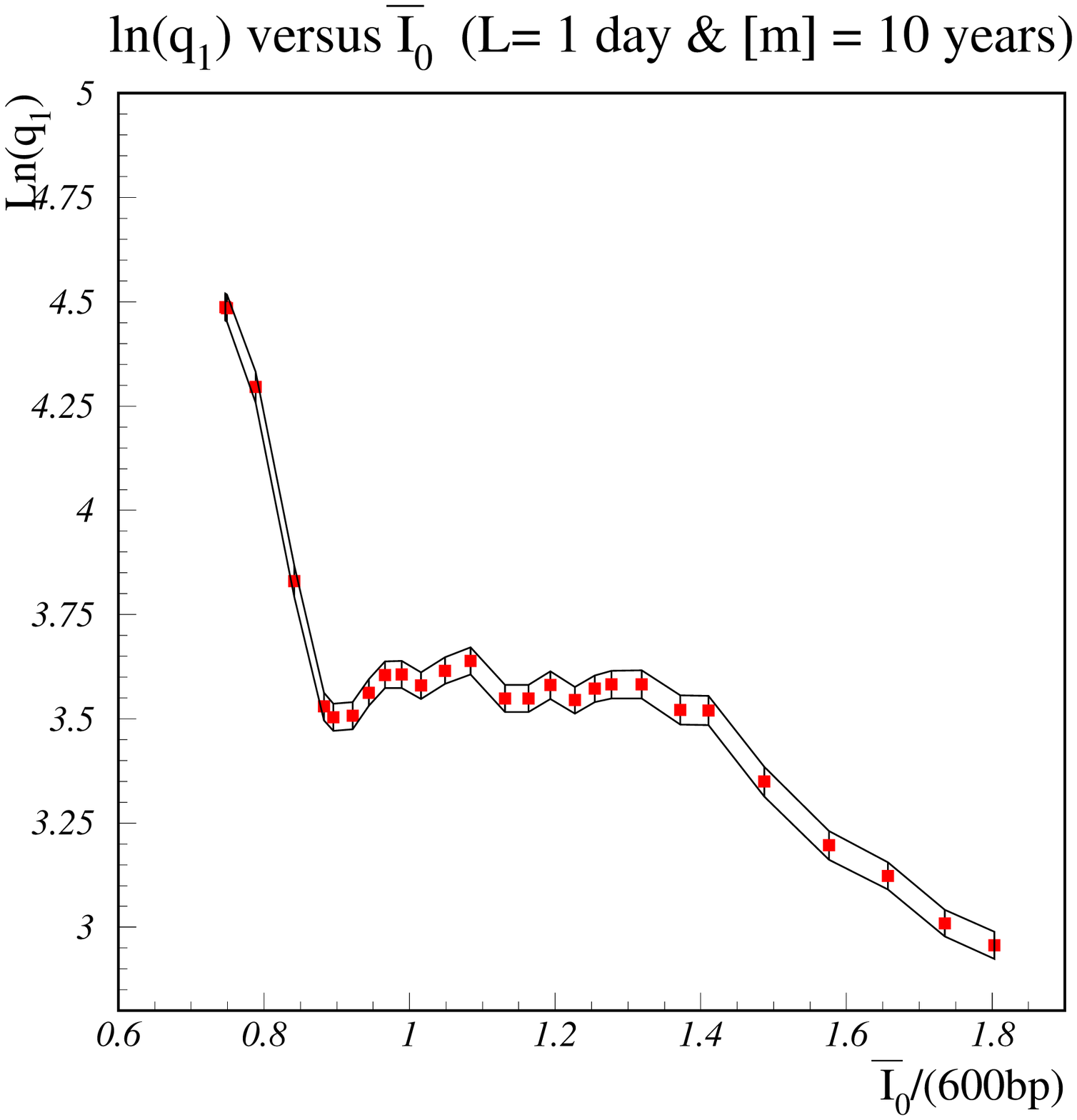}
 &
 \epsfxsize=6.4cm  
 \epsffile{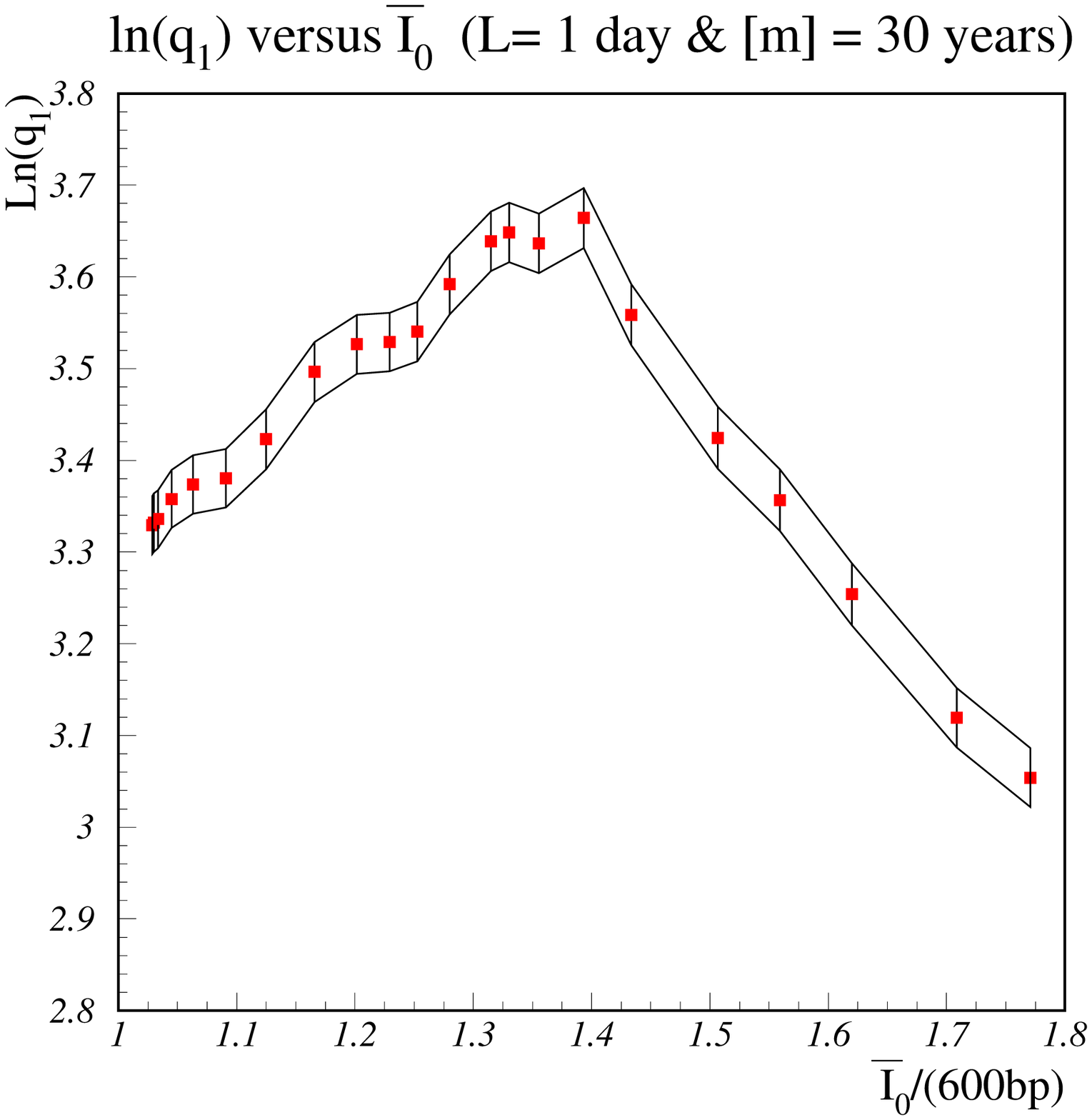}
\\
\end{tabular}
\end{center}

\newpage

\centerline{\bf{Figure \re{q2par2figrest}}}
\vspace{-0.7cm}
\begin{center}
\begin{tabular}{cc}
 \epsfxsize=6.4cm  
 \epsffile{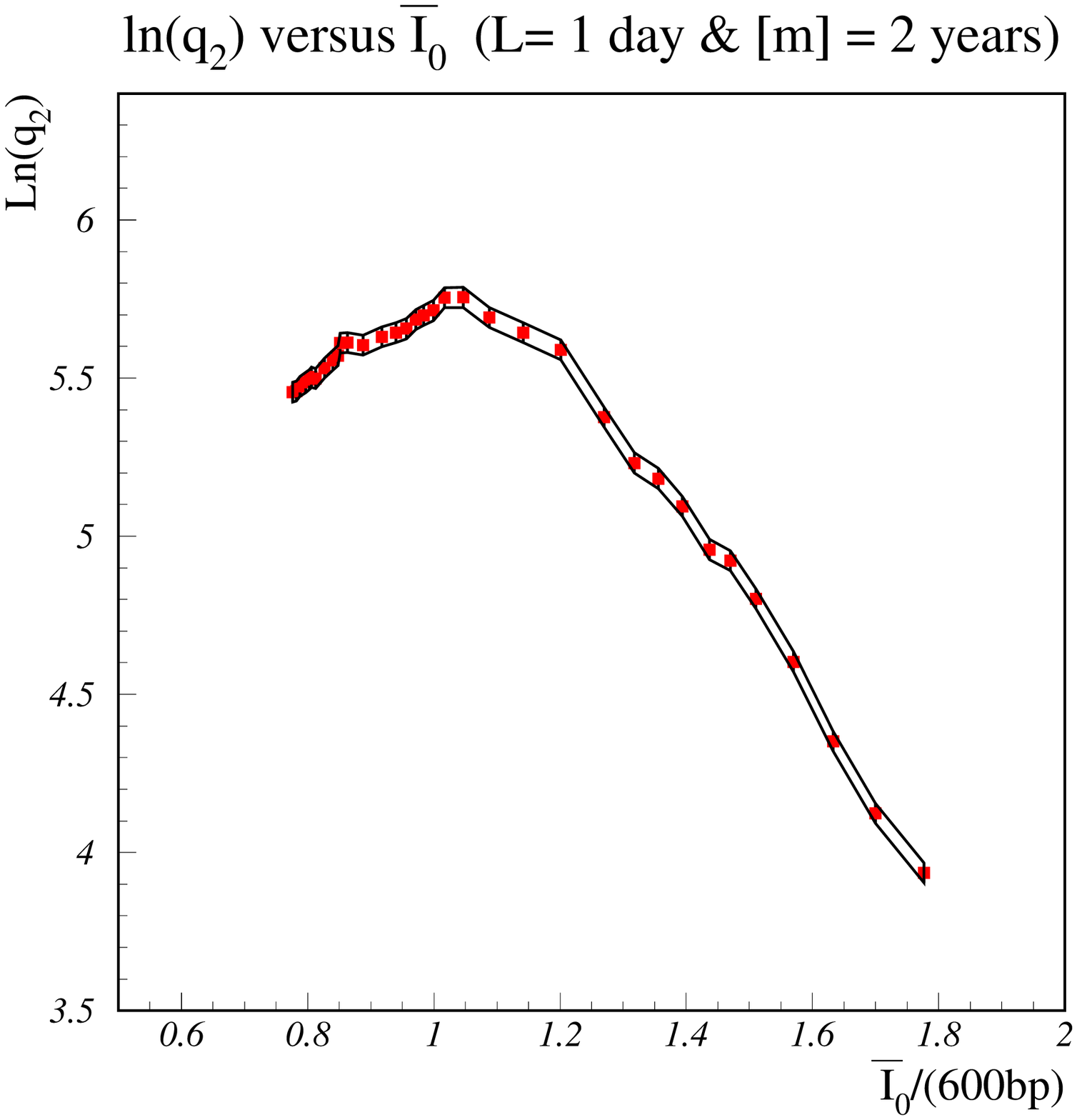}
 &
 \epsfxsize=6.4cm  
 \epsffile{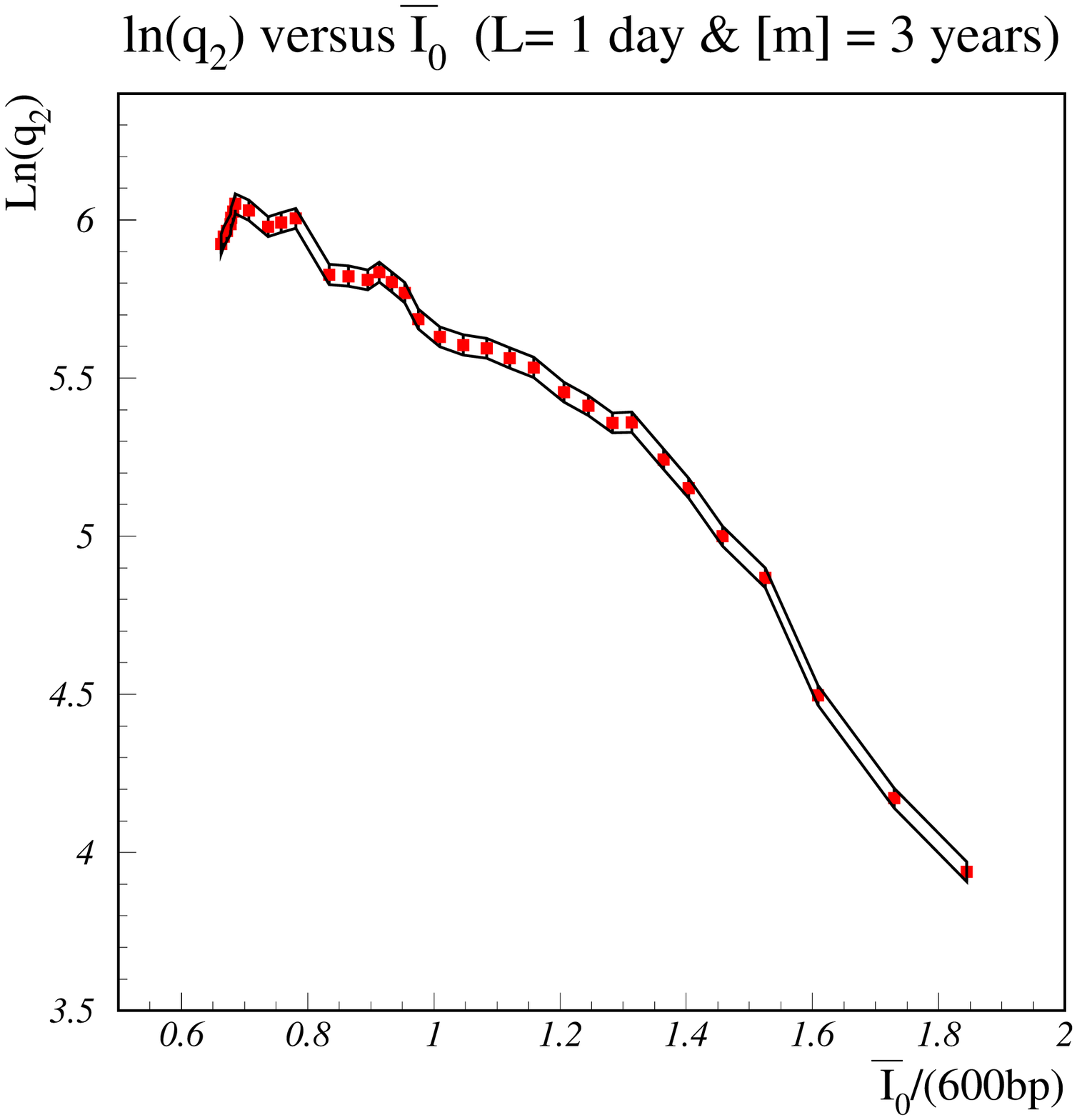}
 \\
 \epsfxsize=6.4cm  
 \epsffile{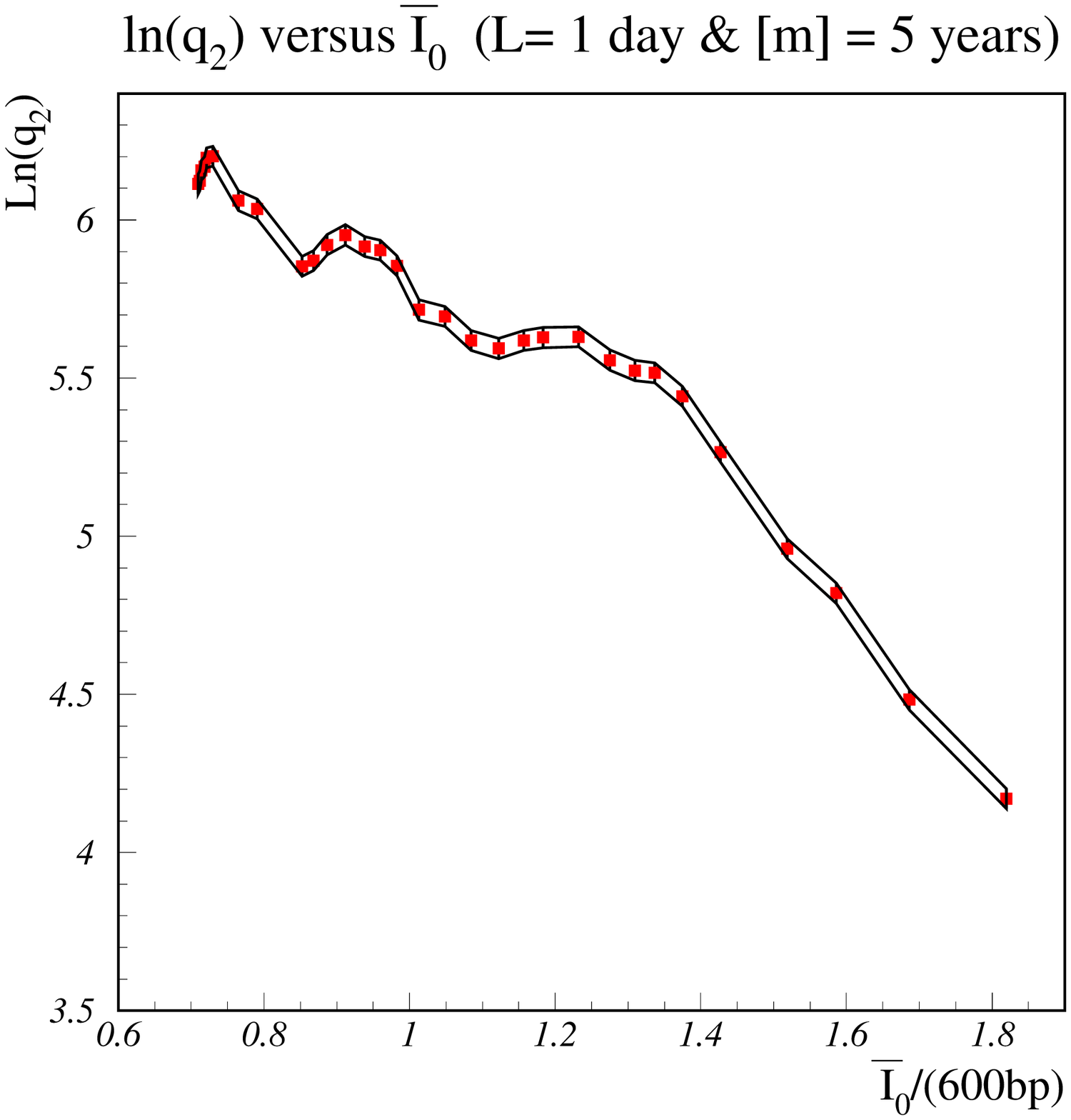}
 &
 \epsfxsize=6.4cm  
 \epsffile{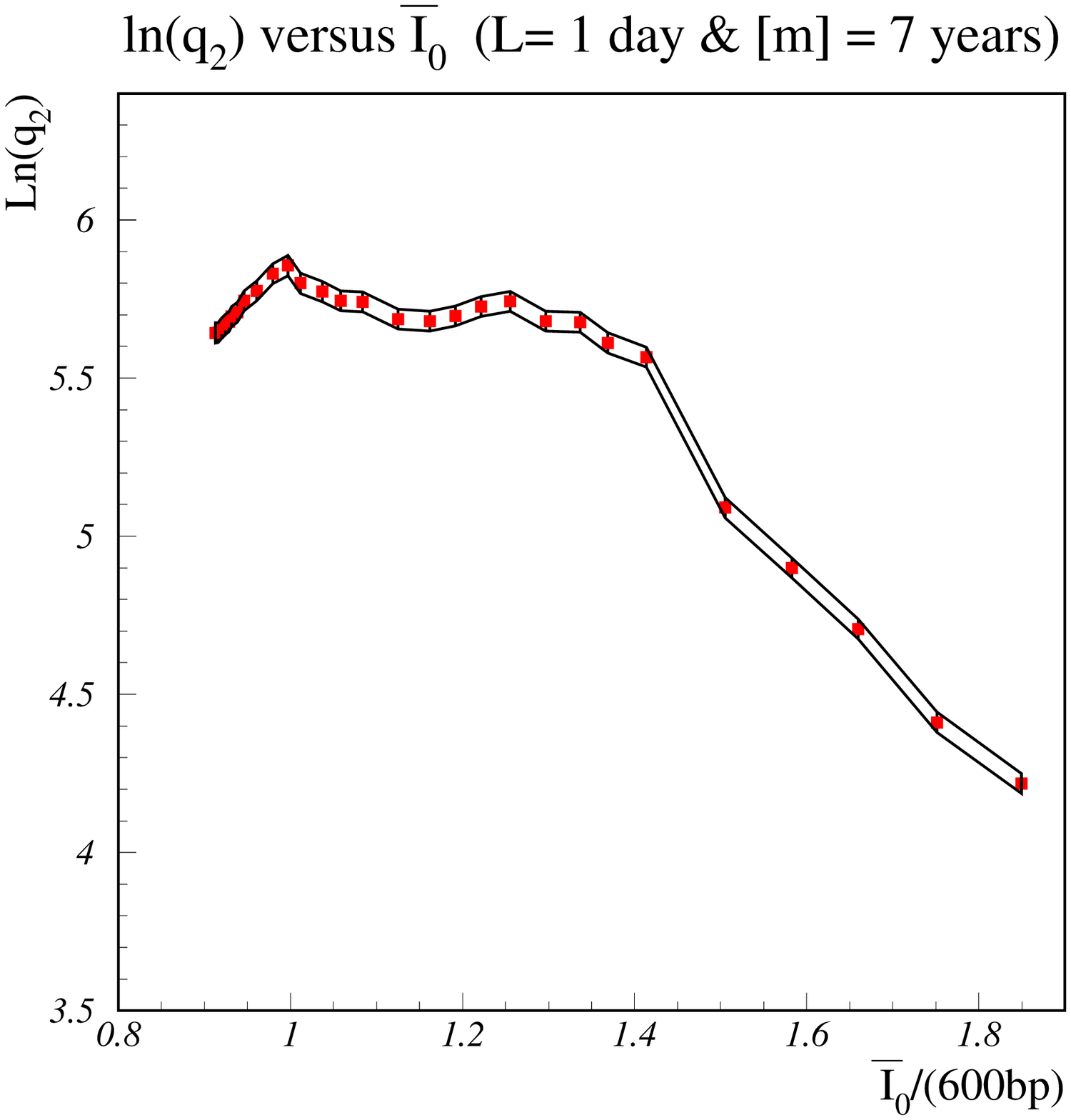}
 \\
 \epsfxsize=6.4cm  
 \epsffile{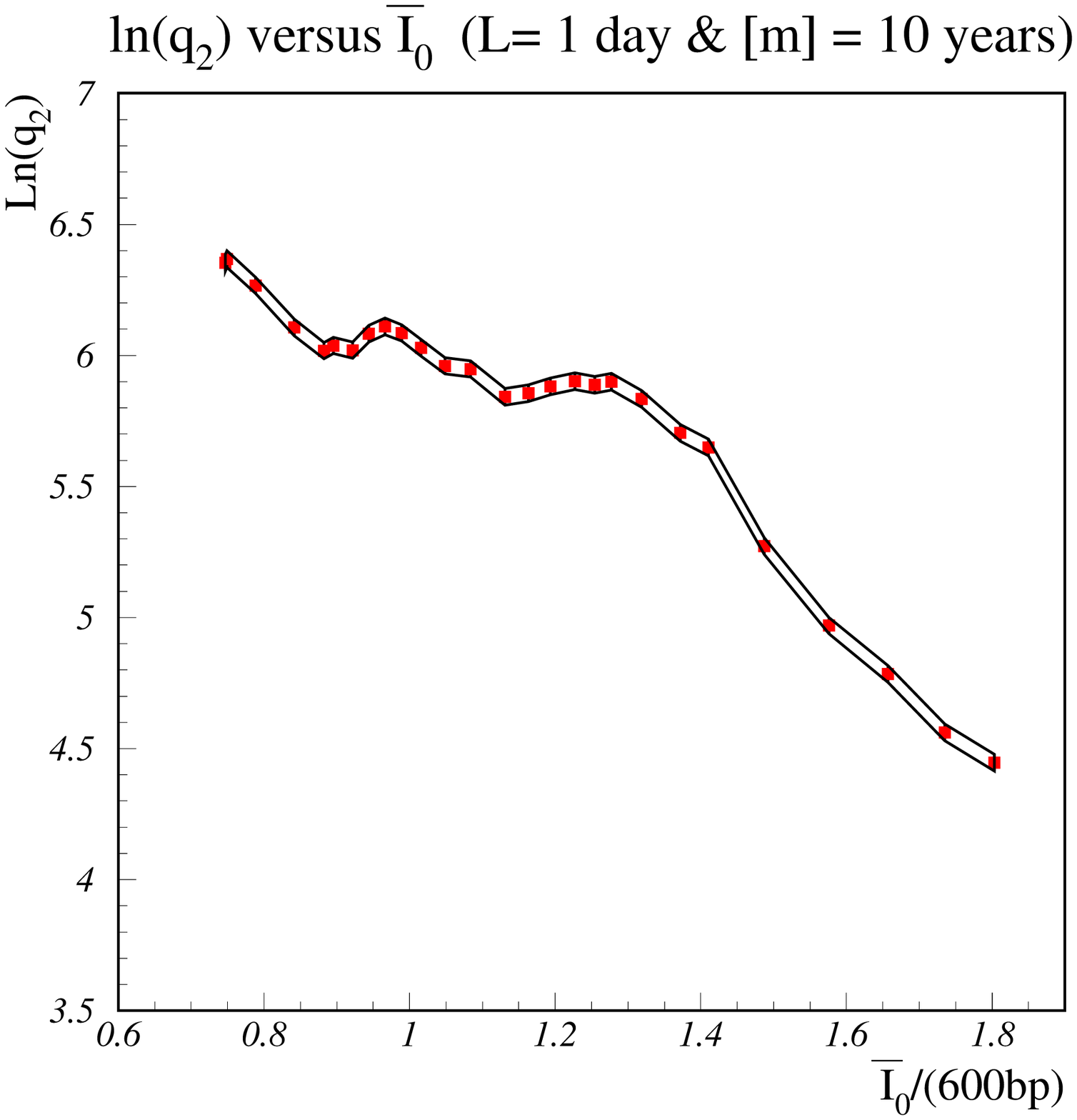}
 &
 \epsfxsize=6.4cm  
 \epsffile{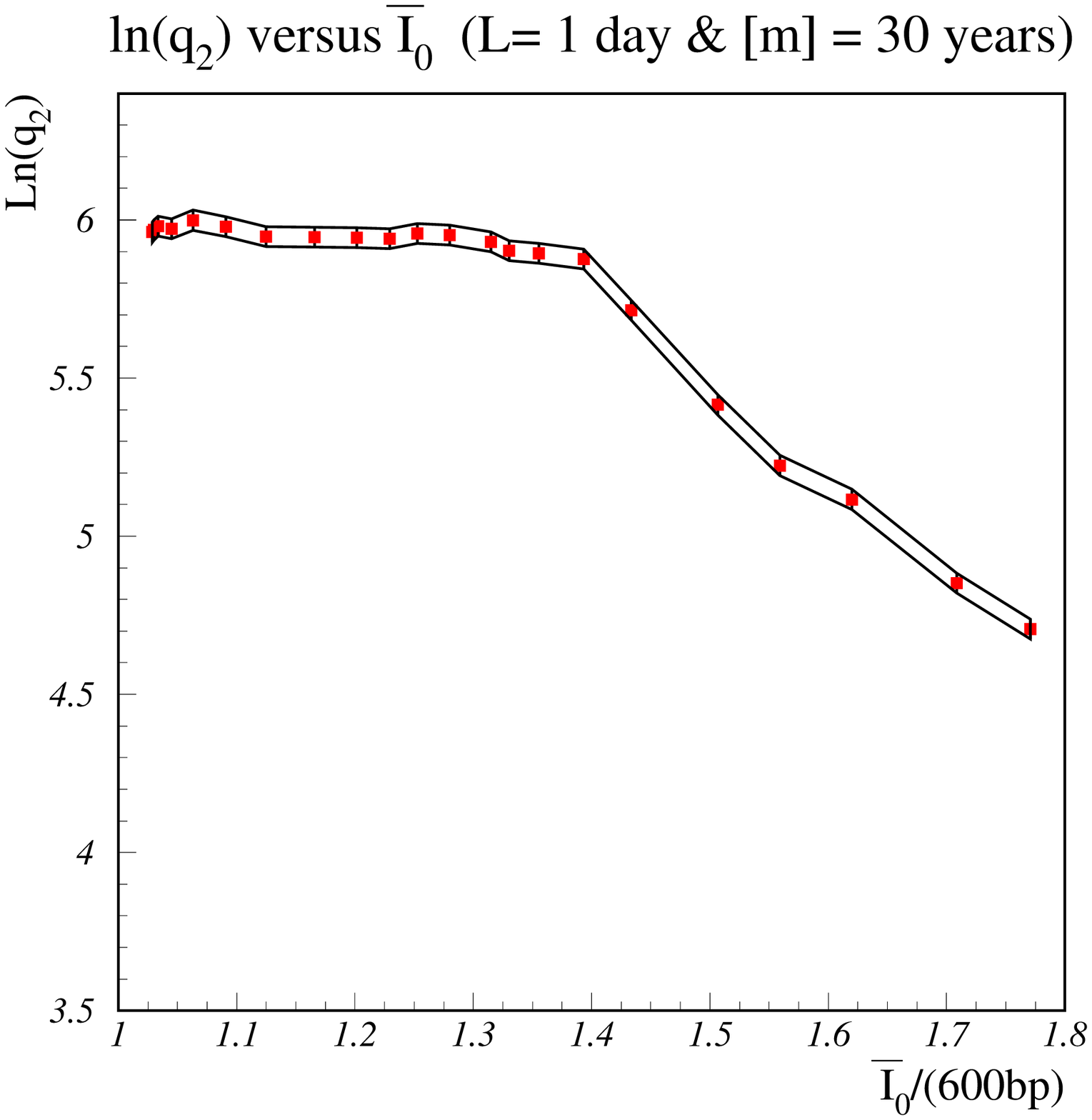}
\\
\end{tabular}
\end{center}

\newpage
\centerline{\bf{Figure \re{deltaq1}}}
\begin{center}
\epsfxsize=8.5cm\epsffile{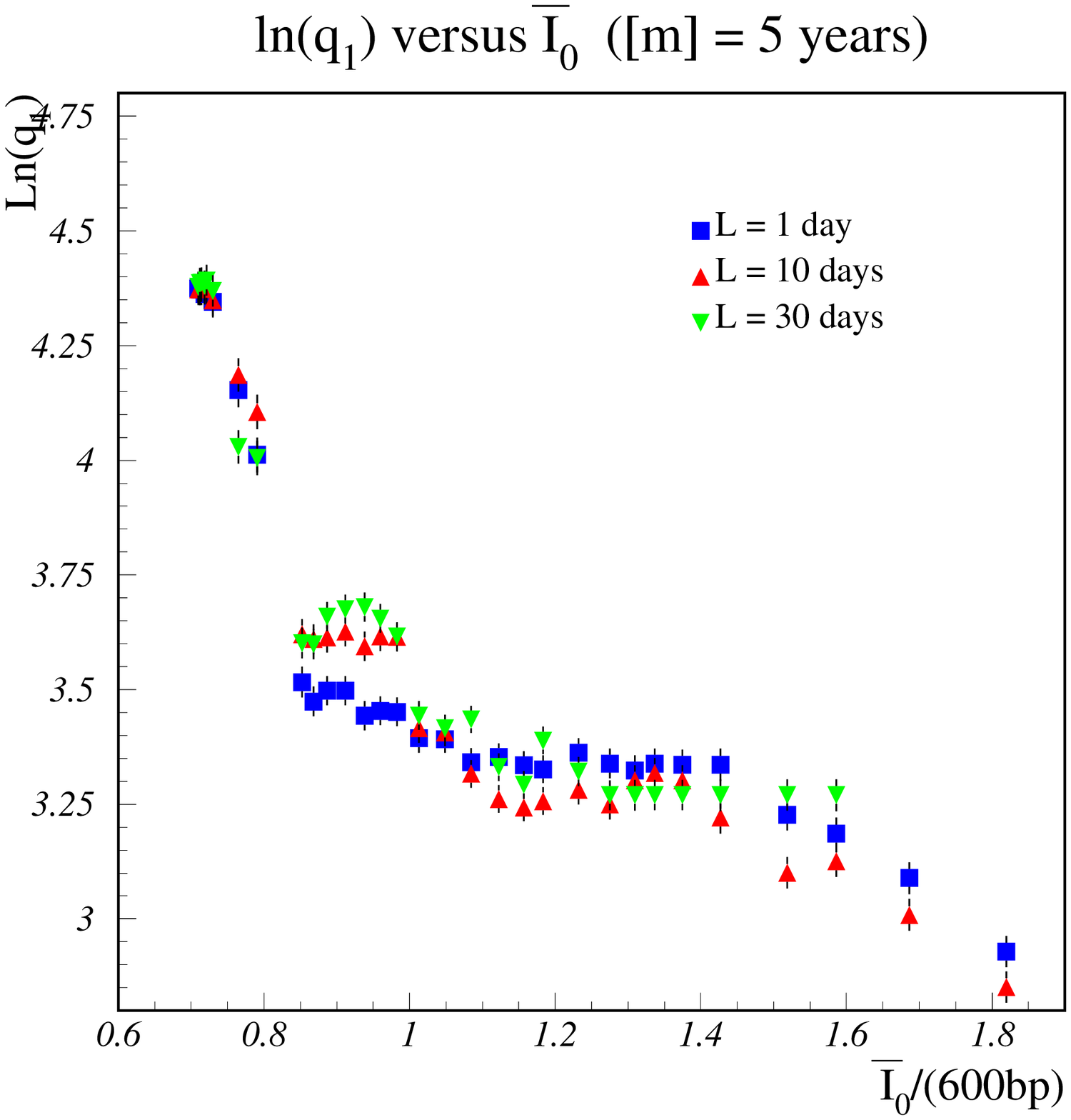}
\end{center}

\centerline{\bf{Figure \re{deltaq2}}}
\begin{center}
\epsfxsize=8.5cm\epsffile{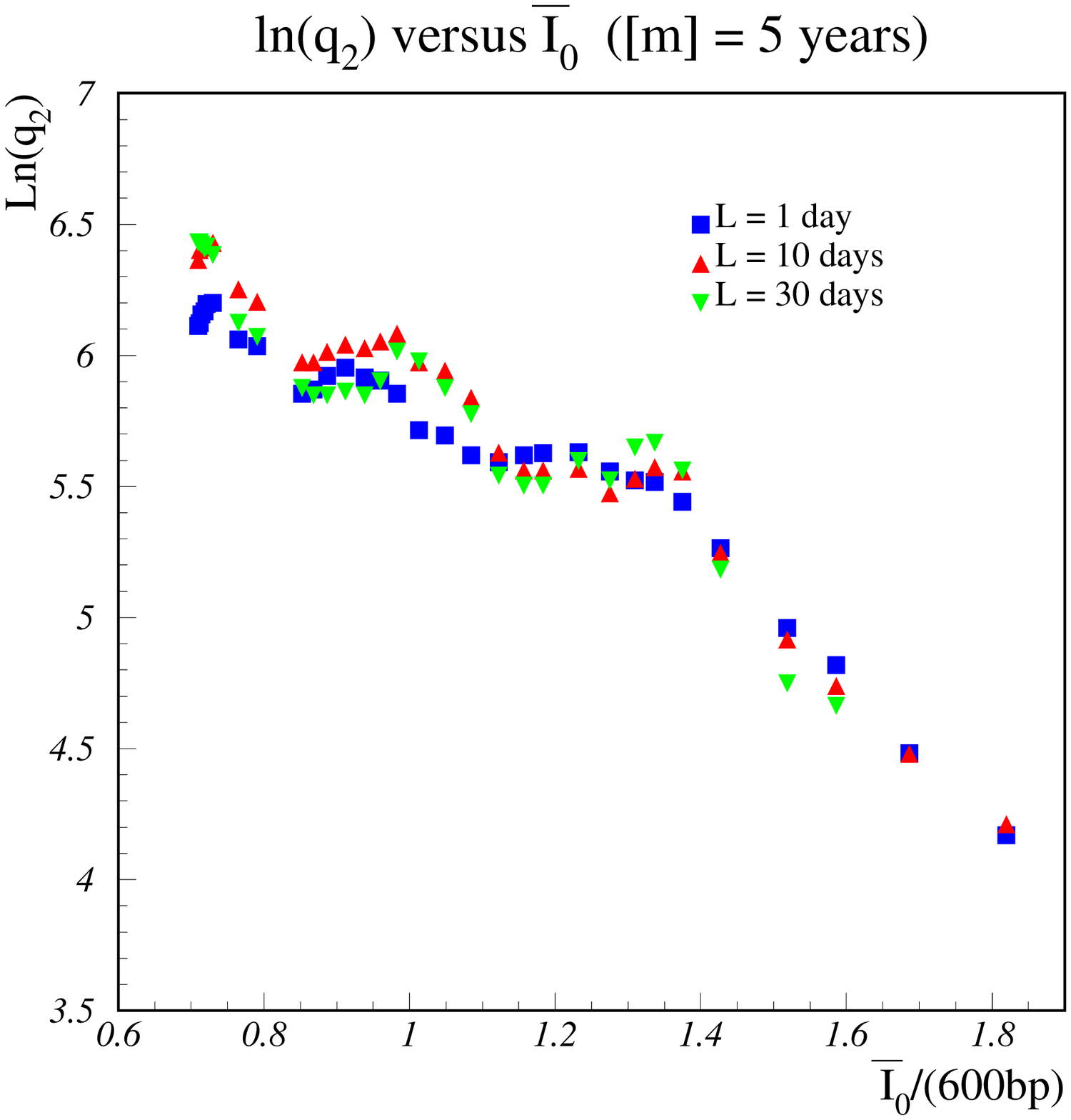}
\end{center}

\newpage

\centerline{\bf{Table \re{mui}}}
\begin{center}
Data set \{2\} \\
~\\
\begin{tabular}{c c c c}
\hline
maturity (yrs) & $\mu_1$ (\%$^{-1}$)  & $\mu_2$ (\%$^{-2}$) & ${\mu_2}/{\mu_1}$ (\%$^{-1}$) \\
\hline
1 & 0.657 $\pm$ 0.005 & 1.197 $\pm$ 0.005 & 1.82 $\pm$ 0.02 \\
2 & 0.627 $\pm$ 0.006 & 1.178 $\pm$ 0.007 & 1.88 $\pm$ 0.03 \\
3 & 0.653 $\pm$ 0.005 & 1.164 $\pm$ 0.006 & 1.78 $\pm$ 0.02 \\ 
5 & 0.638 $\pm$ 0.005 & 1.139 $\pm$ 0.005 & 1.79 $\pm$ 0.02  \\ 
7 & 0.626 $\pm$ 0.005 & 1.111 $\pm$ 0.005 & 1.78 $\pm$ 0.02  \\ 
10 & 0.636 $\pm$ 0.005 & 1.100 $\pm$ 0.005 & 1.73 $\pm$ 0.02  \\
30 & 0.578 $\pm$ 0.006 & 1.068 $\pm$ 0.006 & 1.85 $\pm$ 0.02 \\
\hline
\end{tabular}
\end{center}

\centerline{\bf{Table \re{muiold}}}
\begin{center}
Data set \{1\} \\
~\\
\begin{tabular}{c c c c}
\hline
maturity (yrs) & $\mu_1$ (\%$^{-1}$)  & $\mu_2$ (\%$^{-2}$) & ${\mu_2}/{\mu_1}$ (\%$^{-1}$) \\
\hline
1 & 0.607 $\pm$ 0.006 & 1.209 $\pm$ 0.009 & 1.99 $\pm$ 0.03  \\
2 & 0.644 $\pm$ 0.004 & 1.200 $\pm$ 0.007 & 1.86 $\pm$ 0.03 \\
3 & 0.652 $\pm$ 0.004 & 1.172 $\pm$ 0.007 & 1.80 $\pm$ 0.02 \\ 
5 & 0.622 $\pm$ 0.006 & 1.139 $\pm$ 0.005 & 1.83 $\pm$ 0.02 \\ 
7 & 0.606 $\pm$ 0.006 & 1.109 $\pm$ 0.004 & 1.83 $\pm$ 0.02 \\ 
10 & 0.597 $\pm$ 0.005 & 1.093 $\pm$ 0.005 & 1.83 $\pm$  0.02 \\
30 & 0.588 $\pm$ 0.005 & 1.074 $\pm$ 0.004 & 1.83  $\pm$ 0.02 \\
\hline
\end{tabular}
\end{center}

\end{document}